\documentclass[11pt,preprint]{aastex}
\usepackage{epsf}

\newcommand{\hubunits}{\;{\rm km}\,{\rm s}^{-1}\,{\rm Mpc}^{-1}}

\newcommand{\hmpc}{h^{-1}{\rm Mpc}}

\newcommand{\Mbmin}{M_{b,{\rm min}}}
\newcommand{\Mhmin}{M_{h,{\rm min}}}
\newcommand{\Msmin}{M_{s,{\rm min}}}
\newcommand{\Navg}{\langle N \rangle_M}
\newcommand{\NNavg}{\langle N(N-1) \rangle_M}
\newcommand{\NNNavg}{\langle N(N-1)(N-2) \rangle_M}
\newcommand{\vunits}{h^3{\rm Mpc}^{-3}}
\newcommand{\adapt}{{\tt AdaptaHOP}}

\slugcomment{Submitted to ApJ}

\shortauthors{Weinberg et al.}
\shorttitle{Baryons and Dark Matter}
\singlespace

\begin{document}

\title{Baryon Dynamics, Dark Matter Substructure, and Galaxies}

\author{David H. Weinberg}
\affil{Astronomy Department, Ohio State University, Columbus, OH 43210;
dhw@astronomy.ohio-state.edu}
\author{Stephane Colombi}
\affil{Institut d'Astrophysique de Paris, 75014 Paris, France;
colombi@iap.fr}
\author{Romeel Dav\'e}
\affil{Steward Observatory, University of Arizona, Tucson, AZ 85721;
rad@as.arizona.edu}            
\author{Neal Katz}
\affil{Astronomy Department, University of Massachusetts, Amherst, MA 01003;
nsk@kaka.astro.umass.edu}

\begin{abstract}
By comparing a collisionless cosmological N-body simulation (DM) to a smoothed
particle hydrodynamics simulation (SPH) with the same initial conditions, we 
investigate the correspondence between the dark matter subhalos produced by
collisionless dynamics and the galaxies produced by dissipative gas 
dynamics in a dark matter background. When galaxies in the SPH simulation fall
into larger groups and become satellites, they retain local dark matter
concentrations (SPH subhalos) whose mass is typically five times the galaxy
baryonic mass (compared to the simulation's universal ratio 
$\Omega_{\rm dm}/\Omega_b \approx 7.5$). The more massive subhalos of the
SPH simulation generally have corresponding subhalos of similar mass and
spatial position in the DM simulation; at lower masses, there is still fairly
good correspondence, but some DM subhalos are in different spatial positions
and some have suffered tidal stripping or disruption. The halo occupation 
statistics of DM subhalos --- the mean number of subhalos, pairs, and triples 
as a function of host halo mass --- are very similar to those of SPH subhalos 
and SPH galaxies. The gravity of the dissipative baryon component amplifies the
density contrast of subhalos in the SPH simulation, making them more resistant
to tidal disruption. Relative to SPH galaxies and SPH subhalos, the DM subhalo
population is depleted in the densest regions of the most massive halos. The
good agreement of halo occupation statistics between the DM subhalo and SPH
galaxy populations leads to good agreement of their two-point correlation 
functions and higher order moments on large scales. The depletion of DM
subhalos in dense regions depresses their clustering at $R <1\hmpc$. In these
simulations, the ``conversation'' between dark matter and baryons is mostly
one-way, with dark matter dynamics telling galaxies where to form and how to
cluster, but the ``back talk'' of the baryons influences small scale clustering
by enhancing the survival of substructure in the densest environments.
\end{abstract}


\section{Introduction}
\label{sec:intro}

The idea that galaxies form by dissipation of the baryonic component
within a collisionless dark matter halo has a long history \citep{white78}.
The excellent agreement of the inflationary cold dark matter (CDM)
model with a wide range of cosmological observations 
(e.g., \citealt{spergel03,seljak05}) puts this hypothesis on a firm theoretical 
footing.  In the first stages of galaxy formation, dark matter does the
talking: gravitational collapse produces dark matter potential wells
that capture baryons, which radiate their energy and form dense 
objects at the halo centers.  However, the subsequent details of the 
baryon-dark matter ``conversation'' are not so clear.
Early N-body simulations showed that mergers of dark matter halos
were followed by fairly rapid erasure of substructure,
suggesting that dissipation in the baryonic component was crucial
to the formation of groups and clusters with many distinct members,
and that it was the gravity of the condensed baryons that allowed
them to retain the central regions of their individual
dark matter halos after falling into larger virialized systems.
However, much higher resolution simulations in the late 1990s and
thereafter showed
that virialized halos retain a great deal of substructure 
\citep{ghigna98,klypin99,moore99,springel01}, 
and that the erasure of substructure in earlier 
simulations was largely an artifact of inadequate mass and/or force 
resolution.  This result raises the possibility that baryon self-gravity
is unimportant in producing groups and clusters,
and that cooling and star formation merely produce ``beacons'' that mark
the centers of dark matter structures that would be present even if 
baryons had no gravitational influence at all.

In this paper, we investigate the dynamical interaction between 
baryons and dark matter using two numerical simulations
of the same cosmological volume, one that incorporates both dark matter
and a baryonic component modeled with smoothed particle hydrodynamics (SPH),
and the other that starts from the same initial conditions but follows only
the dark matter.  We identify substructure in the dark matter distribution
using a method that computes SPH-like density estimates within halos,
then groups particles above saddle points in the density field.
We are interested in the degree to which the presence of baryons alters
the properties of substructure in the dark matter distribution and in
the degree to which substructure in the purely gravitational simulation
traces the galaxy population that forms in the hydrodynamic simulation.
The latter issue is of practical as well of physical interest,
since if the agreement is good one might be able to use N-body simulations
in place of hydrodynamic simulations for galaxy clustering predictions.
N-body simulations with the resolution needed to follow substructure
are computationally expensive, but they are less expensive than full
hydrodynamic simulations, thus allowing larger simulation volumes or 
wider searches of parameter space.
We therefore pay particular attention to the halo occupation statistics
of substructures vs. galaxies, since these in turn allow one to predict
many different clustering statistics \citep{berlind02}.

There are several indirect indications that substructure
in N-body simulations can provide good tracers of the galaxy population.
First, \cite{colin99} and \cite{kravtsov04} show that
the correlation functions of substructures in high resolution
simulations agree quite well with observations, which in turn agree
well with results from full hydrodynamic simulations 
\citep{cen00,pearce01,yoshikawa01,weinberg04}.
\cite{conroy06} show that the agreement with observations extends
to a wide range of redshifts and galaxy space densities.
Second, \cite{berlind03} show that 
there is remarkably good agreement between the 
halo occupation distribution found in our SPH simulations and those
predicted by semi-analytic galaxy formation models of \cite{cole00}.
Since the treatment of cooling and star formation is quite different
in the two methods, this agreement suggests that the halo occupation
distribution is determined in large part by dark matter dynamics
and halo merger histories, though even if true this does not
guarantee that post-merger substructure will retain the information
about the galaxy population.  Third, \cite{kravtsov04} find
that the halo occupation distribution of substructure in their high 
resolution N-body simulations is similar to that found in our SPH
simulations by \cite{berlind03} and \cite{zheng05}.
While all of these results provide
useful insight into the relative importance of dark matter dynamics
and baryon dissipation in determining the spatial distribution of
galaxies, this paper is the first,
to our knowledge, to carry out the direct test of comparing galaxy
populations in a hydrodynamic simulation of a cosmological volume
to dark matter substructure
in an N-body simulation with the same initial conditions.
\cite{nagai05} have recently carried out a complementary experiment
in simulations of galaxy clusters.

\section{Simulations}
\label{sec:sim}

We analyze two simulations with the same initial conditions, 
one run with dark matter only and one that incorporates a dissipative
gas component and star formation.  We hereafter refer to these
as the DM and SPH simulations, respectively.
The SPH simulation uses a parallel implementation \citep{dave97}
of TreeSPH \citep{hernquist89,katz96} to follow the evolution of
$128^3$ dark matter particles and $128^3$ gas particles 
in a $22.222\hmpc$ comoving box, from $z=49$ to $z=0$.  
We adopt a $\Lambda$CDM cosmological model
(inflationary cold dark matter with a cosmological constant)
with parameters $\Omega_m=0.4$, $\Omega_\Lambda=0.6$, $\Omega_b=0.02h^{-2}$,
$h \equiv H_0/100\hubunits = 0.65$, $n=0.95$, and $\sigma_8=0.8$.
Our choices of $\sigma_8,$ $H_0,$ $n$, and $\Omega_b$ are reasonably close to
the recent estimates from cosmic microwave
background anisotropies and large scale structure data
(e.g., \citealt{spergel03,tegmark04,sanchez06}), 
while our value of $\Omega_m$ is somewhat high.
The dark matter particle mass is
$7.9\times 10^8 M_\odot$, and the SPH particle 
mass is $1.05 \times 10^8 M_\odot$.
The gravitational force softening is a 
$5h^{-1}$ comoving kpc cubic spline, roughly equivalent
to a Plummer force softening of $3.5h^{-1}$ comoving kpc.
The DM simulation uses the same simulation code and the same initial
positions and velocities of dark matter particles.  It has the same
numerical parameters, except that the dark matter particle mass
is increased by a factor of $\Omega_m/(\Omega_m-\Omega_b)$ to 
$8.9\times 10^8 M_\odot$.

Although the volume is much smaller, the mass resolution of the DM simulation
is similar to that of the simulations used by \cite{colin99}
and \cite{kravtsov04} to study the large scale clustering of subhalos
(somewhat higher than their $\Lambda{\rm CDM}_{60}$ run with 
$256^3$ particles in a $60\hmpc$ box and somewhat lower than their
$\Lambda{\rm CDM}_{80}$ run with $512^3$ particles in an $80\hmpc$ box),
and it is slightly higher than that of the recent ``Millennium Run''
simulation (\citealt{springel05}; $2160^3$ particles in a $500\hmpc$ box).
It is not as high as the mass resolution in recent simulations
focused on the substructure distribution in clusters
(e.g., \citealt{diemand04,gao04,nagai05}).
By chance, our simulation forms one
halo that is unusually large for a $22.222\hmpc$ box given
our cosmological parameters, with a mass of $4\times 10^{14} M_\odot$,
allowing us to investigate substructure survival and baryonic
influence in a Virgo-mass galaxy cluster
(see Figure~\ref{fig:sub1} below).  This halo contains about
$4.7 \times 10^5$ dark matter particles within its virial region.
The next most massive halos have masses of $\sim 3\times 10^{13} M_\odot$.

Details of the treatment of radiative cooling, star formation, and galaxy
identification in the SPH simulation
can be found in \cite{katz96}.  In brief, 
gas can dissipate energy via Compton cooling and radiative cooling,
computed assuming primordial composition and the photoionizing
background field of \cite{haardt96}.
Star formation occurs in regions that are Jeans unstable, above
a threshold density ($n_H > 0.1 {\rm cm}^{-3}$), and below a
threshold temperature ($T \leq 30,000\,$K).  
We add the thermal energy from
supernova feedback but it has relatively little impact,
because it is usually deposited in a dense medium with a
short cooling time.  We identify galaxies using the 
Spline Kernel Interpolative DENMAX (SKID)\footnote{We use the 
implementation of SKID by J. Stadel and
T. Quinn, which is publicly available at
{\tt http://www-hpcc.astro.washington.edu/tools/skid.html}.}
algorithm \citep{gelb94,katz96}, which identifies gravitationally
bound clumps of stars and cold ($T \leq 30,000\,$K), 
dense ($\rho_g/\bar{\rho}_g \geq 1000$) gas that
are associated with a common density maximum.
Comparisons among simulations with different resolution show that
the locations and baryonic masses (stars plus cold, dense gas) of
SKID galaxies are robust when the mass exceeds that of about 64 SPH
particles, or $6.8\times 10^9 M_\odot$.  For our lowest mass threshold
sample in this paper, we take a slightly larger minimum mass of 
$7.1 \times 10^9 M_\odot$.  There are 1103 galaxies in the simulation
volume above this mass, making the mean space density of this
sample $0.1\vunits$.  We also consider samples with higher minimum
masses and lower mean space densities, as discussed in \S\ref{sec:results}
below.

\section{Identification of Substructure}
\label{sec:substructure}

Our method of identifying halos and substructures uses the publicly
available code \adapt\ \citep{aubert04}, which
derives from the HOP algorithm of \cite{eisenstein98}.
It is very similar to the
method used by \cite{springel01}, though the implementation here
is independent.
The algorithm is detailed in \cite{aubert04}, so we give only
a brief summary of it here.
We first calculate densities
around each dark matter particle using an SPH-like kernel estimator,
with a cubic spline kernel containing 32 neighbors (just as in the
SPH simulation itself). During that operation, we store as well
the $N_{\rm hop}$ nearest neighbors of each particle, with
$N_{\rm hop}=16$ as advocated by \citet{eisenstein98}.
Then, we partition the ensemble of particles into ``peak patches''.
A peak patch is a set of particles with the same local
density maximum, identified by connecting
each particle to its densest neighbour
among its $N_{\rm hop}$ closest to track the local density gradient.
The connectivity between the peak patches
is dictated by the saddle points in the density field. These points
are found by locating local maxima in the boundaries between peak
patches. In fact, for each pair of peak patches connected through
at least one saddle point, one needs only the saddle point with highest
density.
Then one is ready to construct an ensemble of trees,
the haloes, and the branches of the trees and
their leaves, the substructures,
each leaf corresponding to a unique local maximum, or
peak patch.

\begin{figure}
\centerline{
\epsfxsize=3.5truein
\epsfbox{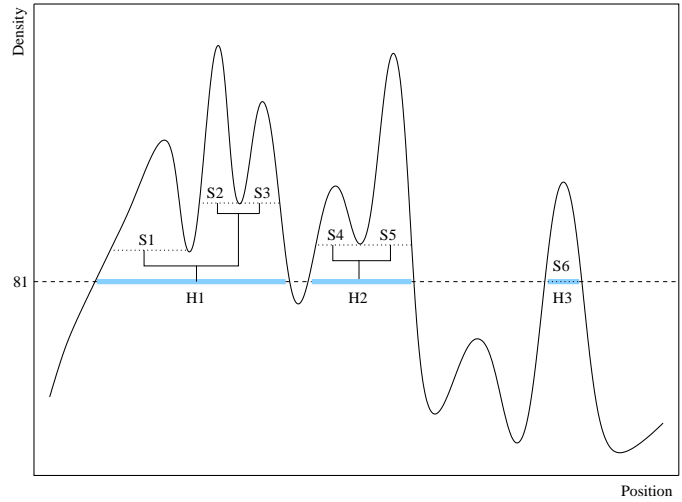}
}
\caption{Illustration of \adapt, in one dimension. In this 
example, saddle points are local minima. The haloes are connected
structures
above the density threshold. Three of them are detected, H1, H2 and H3, as
defined by the thick horizontal lines.
Each elementary substructure (a leaf) is associated to a local maximum.
Therefore, H1, H2 and H3 contain respectively 3, 2 and 1 leaves.
A given substructure can be connected to other ones by saddle points. For
instance S2 is connected to S1 and S3 through 2 saddle points.
To compute the extension of each substructure, we take only the points
which
have density larger than $\rho_{\rm s}$, where $\rho_{\rm s}$ is the
maximum
value measured at the saddle points, as defined by the horizontal dotted
lines.
Note that halo H3 is its own substructure, S6.
Its boundary is defined by the halo density threshold
$\rho/\bar{\rho}=81$.
}
\label{fig:adaptahop}
\end{figure}

In this representation, a halo is defined as a connected group of
particles
with overdensity $\rho/\bar{\rho} > 81$, as advocated by
\cite{eisenstein98};
a leaf is defined as a subset of particles in
a peak-patch with SPH density larger than $\rho_{\rm s}$, where $\rho_{\rm
s}$
is the density of the highest saddle point connecting this peak patch to a
neighbouring
one. In order to select substructures which are statistically significant
compared
to Poisson noise, we impose a 4$\sigma$ level threshold,
\begin{equation}
\langle \rho \rangle_{\rm substructure} > \rho_{\rm s} \left[
1+\frac{4}{\sqrt{N}} \right],
\label{eq:condsad}
\end{equation}
where $N$ is the number of particles in this substructure
(with SPH density above $\rho_{\rm s}$) and
$\langle \rho \rangle_{\rm substructure}$ is the average SPH density
in this substructure. A substructure not following this constraint is
absorbed
by the neighboring substructure connected to it
through the highest saddle point. This operation is
performed recursively until condition (\ref{eq:condsad}) is fullfilled.
Finally, note that most of haloes do not have any substructure, or
equivalently, only one, the halo itself.
In our representation, and in what follows,
such a halo is considered simultaneously as a halo and a
substructure.
A higher resolution simulation would presumably reveal
substructures in these low mass halos, but they would be
below the mass threshold of the resolved galaxy populations
that we consider below.

 To understand the procedure followed in \adapt, it is instructive
to look at the 1-dimensional analogue shown in Figure~\ref{fig:adaptahop}.
The densities at the particle locations define a 1-dimensional
density field.  Halos are connected regions above the overdensity
threshold of 81 (horizontal dashed line), and there are three such halos
in Figure~\ref{fig:adaptahop}
(horizontal thick lines).
Saddle-points in 3-d correspond to local minima in this 1-d example,
and the three halos in Figure~\ref{fig:adaptahop}
contain three substructures, two 
substructures, and no substructure,
respectively.  There is necessarily one and only one maximum between
each pair of minima, and this maximum is identified as the location
of the substructure.  The mass of the substructure is the mass
{\it above} the density threshold of the higher minimum, as
indicated by the horizontal dotted lines in Figure~\ref{fig:adaptahop}.
In most cases, this is a reasonable
way of assigning mass, but it tends to underestimate the mass of
a large central object with a much smaller satellite.
Note that the sum of the masses of the substructures in a halo is
generally smaller than the halo mass itself, unless the halo does
not contain any substructure.
In the present paper, we use only the trees (haloes) and leaves
(individual subhalos), but the substructure finder also builds the entire
set of branches using saddle points as connectors.

\section{Results}
\label{sec:results}

\subsection{Formation of a Galaxy Group}
\label{sec:examples}

Figure~\ref{fig:group} illustrates the formation history of a
representative galaxy group.  This group occupies the fourth
most massive halo in the simulation, with a mass of 
$3.1 \times 10^{13} M_\odot$.  The left and middle columns
show the dark matter particle distributions in the DM and SPH
simulations, respectively, with particles color-coded according
to local density estimated with the 32-particle spline kernel
used in the substructure identification.
Specifically, the lower panels show the dark matter particle
distributions in the central
$0.5 \hmpc$ of this halo at $z=0$; the full extent of the
region within the $\rho/\bar{\rho}>81$ surface is about
a factor of two larger.  The top three rows show the 
distributions of the same particles at $z=2$, 1, and 0.5, respectively
(a small number of particles are missed off the bottom of these plots).
At $z=2$, these particles are spread over a region roughly
$5\hmpc$ across (comoving), and many of them are clumped into small halos
aligned along filamentary structures.  
Between $z=2$ and $z=0.5$, the small halos merge into
larger halos, the underlying filamentary network becomes
less evident, and the whole comoving volume shrinks slightly 
in size.

\begin{figure}
\caption{
{\it (Following page.)}
Formation of a galaxy group, in a halo of mass $3.1 \times 10^{13} M_\odot$.
The lower left panel shows the dark matter particles in the central
$0.5 \hmpc$ of this halo at $z=0$, in the DM simulation.
Panels above it show the positions
of the same particles at $z=3$, 1, and 0.5 (top to bottom).
The panel size is different at each redshift; the white bar is always
$0.5 \hmpc$ comoving.  Central panels show the
corresponding dark matter particle distributions in the SPH simulation.
Right hand panels show the distributions of gas and star particles
at the same redshifts.  Green points (plotted larger for visibility at
$z>0$) show star particles.  Gas particles are color coded by temperature
on a scale running from $\sim 5\times 10^3\,$K (blue) to 
$\sim 5\times 10^6\,K$ (red).
}
\label{fig:group}
\end{figure}


\begin{figure}
\centerline{
\epsfxsize=6.2truein
\epsfbox[77 87 535 704]{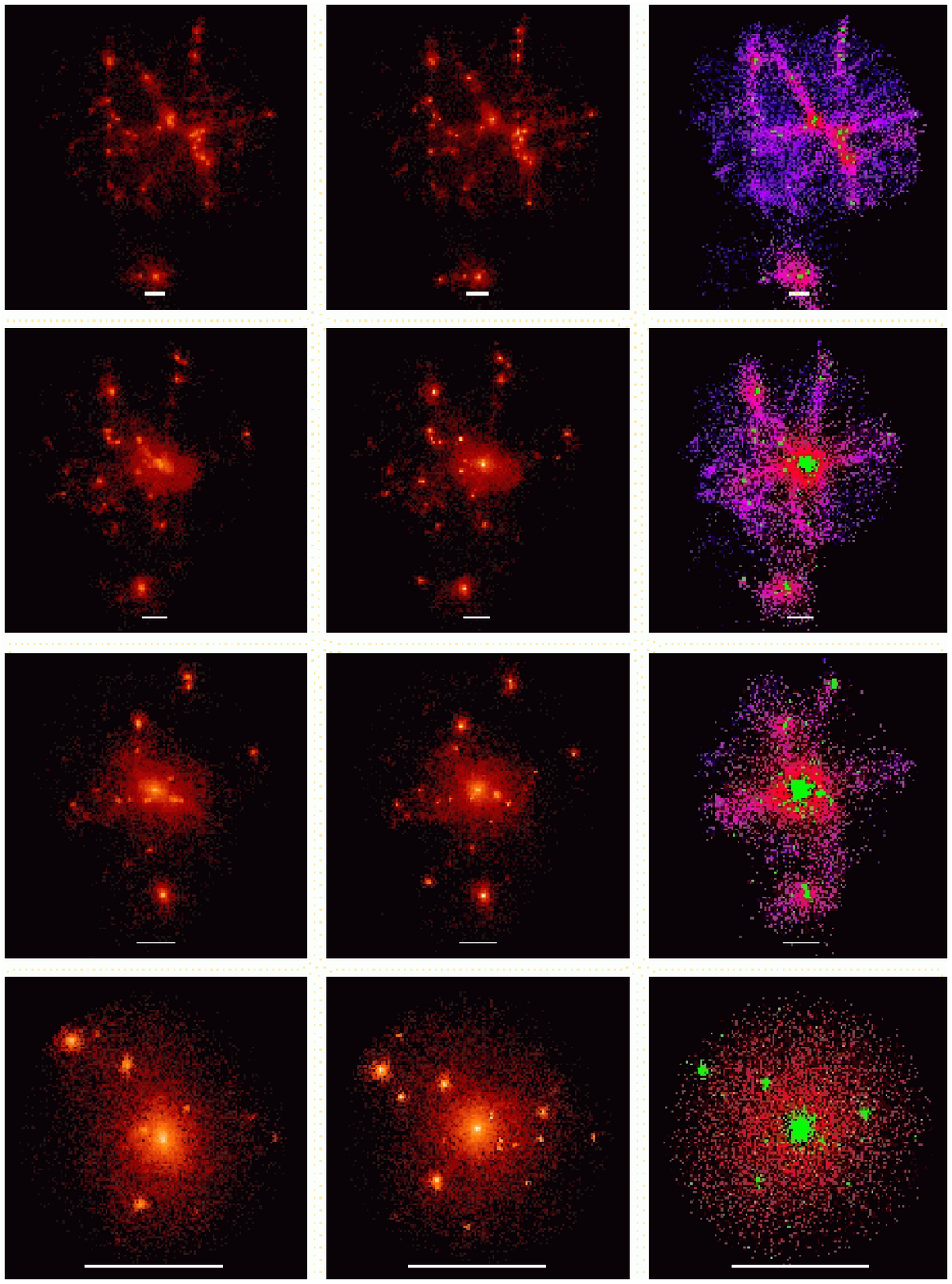}
}
\end{figure}

The right hand column of Figure~\ref{fig:group}
shows the distribution of star particles, shown as green dots,
and gas particles, color coded by temperature, from the SPH simulation.
The temperature color scale runs logarithmically from 
$T \sim 5\times 10^3\,$K (blue) to
$T \sim 10^7\,$K (yellow).
Note that high density can make the stellar clumps visually
inconspicuous even when they are fairly massive; we have used
larger dots for star particles at $z \geq 0.5$ so that the
clumps remain visible.  By $z=0$, all of the gas that is not in
galaxies has been heated to $T \sim 5\times 10^6\,$K, but
at earlier times much of the gas in
filaments or the diffuse medium between them is cooler than 
$10^5\,$K.  The high redshift panels also show clumps of dense
gas that has cooled to $T\sim 10^4\,$K but has not yet
formed stars.  The absence of stars in these clumps is
primarily a numerical resolution effect --- in tests with
simulations of varying resolution, we find that the SPH code
underestimates star formation rates in objects with less 
than $\sim 200$ particles.

At $z\geq 1$, the dark matter distributions in the simulations
with and without gas are nearly indistinguishable, and even
at $z=0.5$, the differences are small.  Gas condensation and
star formation occur at the centers of the larger dark matter
concentrations.  At $z=0$, all of the larger galaxies are
associated with a visually identifiable dark matter substructure.
The largest substructures are at similar locations in the DM and SPH
simulations.  Smaller substructures cannot be visually matched one-to-one
between the two simulations.  The gravity of the dense baryon 
clumps increases the density of subhalos in the SPH simulation,
making them more visually prominent.  However, we will show
below that the number and mass distribution of subhalos is actually
similar in the SPH and DM simulations, and that the level of one-to-one
subhalo correspondence is more than meets the eye.

\subsection{Galaxy and Subhalo Populations}
\label{sec:populations}

\begin{figure}
\centerline{
\epsfxsize=5.5truein
\epsfbox[17 110 592 684]{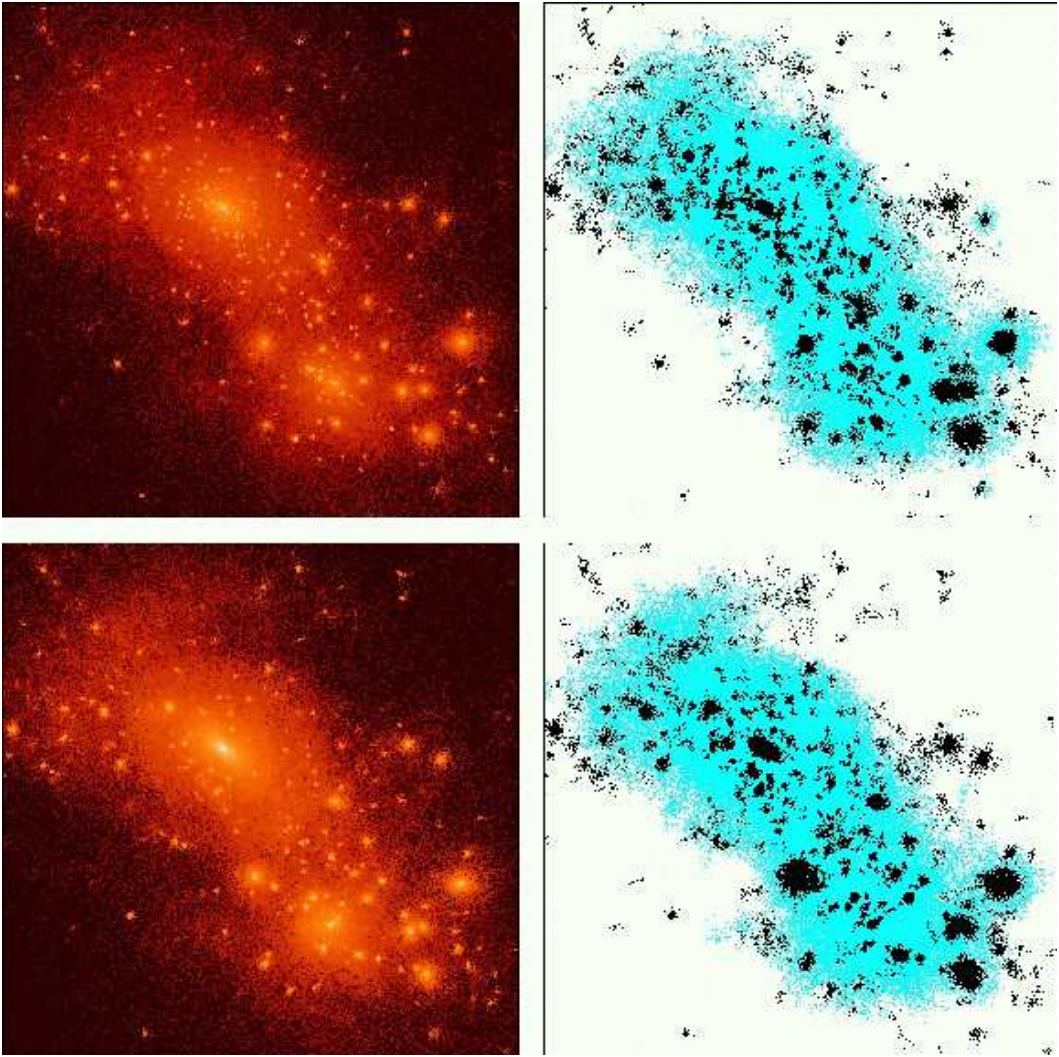}
}
\caption{
Dark matter and dark matter substructure for the largest halo
in the simulation, with a mass of $4.0 \times 10^{14} M_\odot$ at $z=0$.
Panels are $4\hmpc$ across.
Left hand panels show the
halo dark matter particles in the SPH simulation (top)
and DM simulation (bottom).  Particles are coded by local density, 
estimated using a spline kernel enclosing 32 neighbors.
In the right hand panels, black points show
particles that are members
of subhalos identified by \adapt.  Cyan points show particles that
are not connected to one of these substructures but are above the
overdensity 81 threshold.
}
\label{fig:sub1}
\end{figure}

Figure~\ref{fig:sub1} shows the largest halo in the simulation,
with a mass of $4.0\times 10^{14} M_\odot$.
The left hand panels show the dark matter particle distribution
in the SPH simulation (top) and DM simulation (bottom), with
particles color-coded according to their local density, again estimated
with the 32-particle spline-kernel smoothing used in the substructure
identification.  The halo contains
two major subcomponents within its $\rho/\bar{\rho}=81$ 
density boundary.  One can see a large number of local density maxima in 
both the SPH and DM simulations.  There is good correspondence in the
positions of the larger density maxima, while the smaller density peaks
are similar in number but do not correspond in position.
One can also see that these local density peaks are systematically
suppressed in the DM simulation near the centers of the two large clumps. 
A higher resolution simulation might preserve a larger degree of
substructure in these innermost regions, but the SPH and DM 
simulations have the same mass and force resolution, so the
differential effect of including the dissipative baryon
component should be correct, at least qualitatively.
Several groups have recently carried out detailed convergence 
tests for cluster substructure in N-body simulations and concluded
that the suppression of substructure in cluster cores is for the
most part a real effect of tidal stripping and disruption rather
than a numerical artifact \citep{diemand04,gao04}.

The right hand panels of Figure~\ref{fig:sub1} illustrate the application
of \adapt\ to this halo.  Black points show particles that
have been assigned to a subhalo, and cyan points show particles that
are assigned to the parent halo but not to a subhalo.  One can
see a good correspondence between the positions of the largest identified
subhalos in the two simulations, as expected from the density maps.
It is hard to infer the mass of substructures from these plots
because of saturation; denser subhalos often appear less massive
because they are more compact.

\begin{figure}
\centerline{
\epsfxsize=5.5truein
\epsfbox[50 185 450 720]{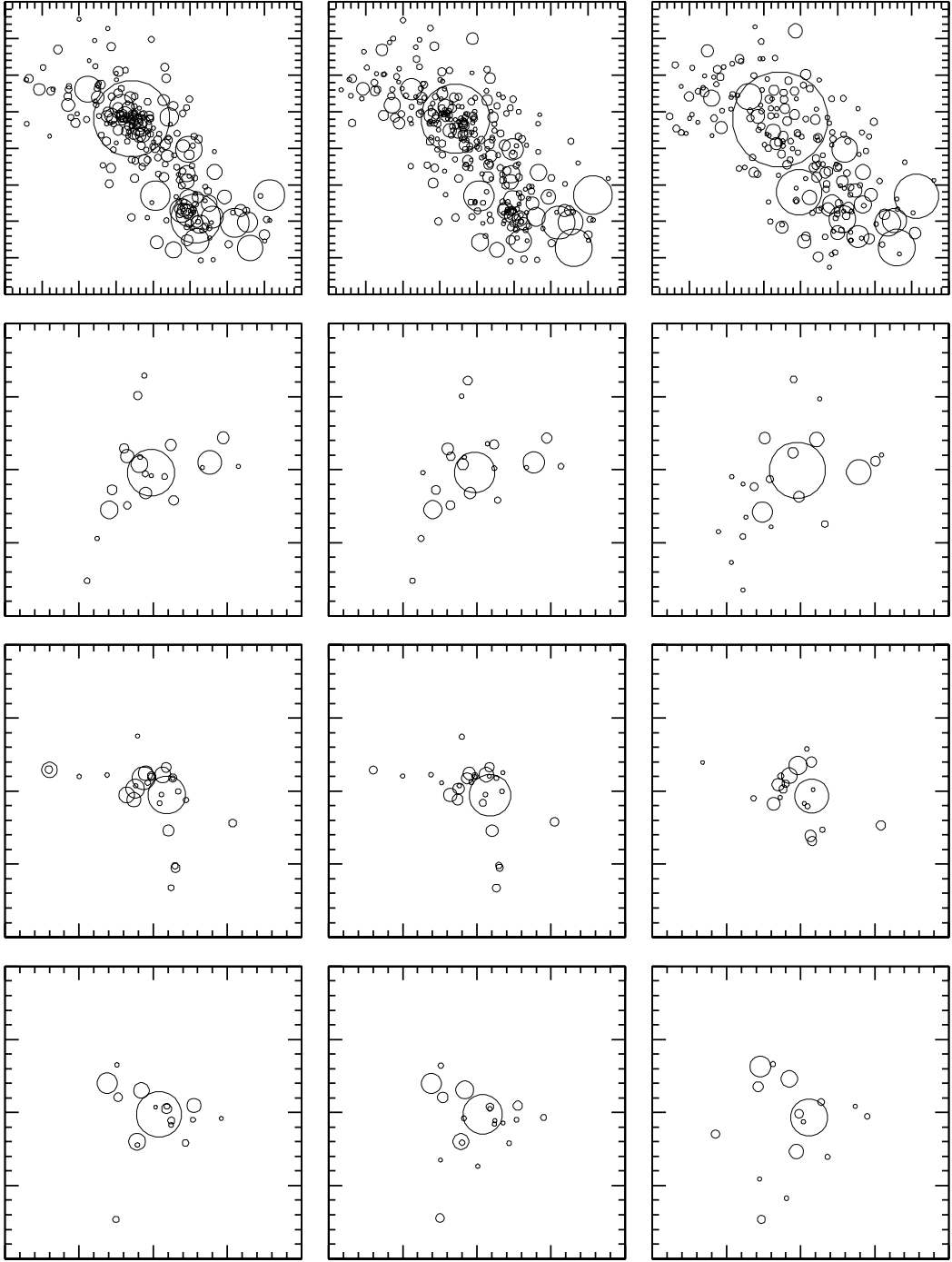}
}
\caption{
Galaxy and subhalo content of the four most massive halos.
Left hand panels show the SPH galaxies, middle panels show the
dark matter subhalos in the SPH simulation, and right panels
show the subhalos in the DM simulation.  Each galaxy or subhalo
is represented by a circle with an area proportional to its mass;
the masses of the subhalos have been multiplied by 0.2 but
are otherwise on the same scale as the galaxies.  The upper panels
are $4\hmpc$ on a side, while the other panels are $2\hmpc$ on a side.
}
\label{fig:circle}
\end{figure}

Figure~\ref{fig:circle} compares the subhalo populations of the two simulations
to the galaxy populations of the SPH simulation, this time for the four
most massive halos.  The bottom row shows the same halo whose
formation history is illustrated in Figure~\ref{fig:group}.
In the left hand panels, each galaxy is represented
by a circle whose area is proportional to its baryonic mass (stars plus
cold gas).  The smallest circles correspond to a mass of 
$7.1\times 10^9 M_\odot$, slightly above our mass resolution limit.
The middle panels show subhalos of the SPH simulation
represented in the same fashion, except that all subhalo masses have
been lowered by a factor of five (and the same point size scaling
and minimum mass threshold have then been applied).
Right hand panels show the DM simulation's subhalo population,
with the same factor of five mass scaling.
The limiting subhalo mass corresponds to 40 dark matter 
particles.\footnote{In this and all subsequent figures, we have 
multiplied the particle masses in the SPH simulation by
$\Omega_m/(\Omega_m-\Omega_b)=1.134,$ so that SPH and DM
subhalos with the same number of {\it particles} are assigned
the same mass.}

There is good agreement between the locations and scaled masses of
the SPH galaxies and the DM subhalos in the SPH simulation, and this
agreement holds almost all the way to the resolution limit except
in the most massive halo.  More remarkably, there is good agreement
between the locations and scaled masses of subhalos in the DM
simulation and the galaxies (and subhalos) in the SPH simulation.
There are some positional differences, and these become larger
for lower mass subhalos, so at low masses it is difficult to 
tell whether there is still a one-to-one correspondence between
subhalos in the two simulations.

\begin{figure}
\centerline{
\epsfxsize=5.5truein
\epsfbox[65 255 560 725]{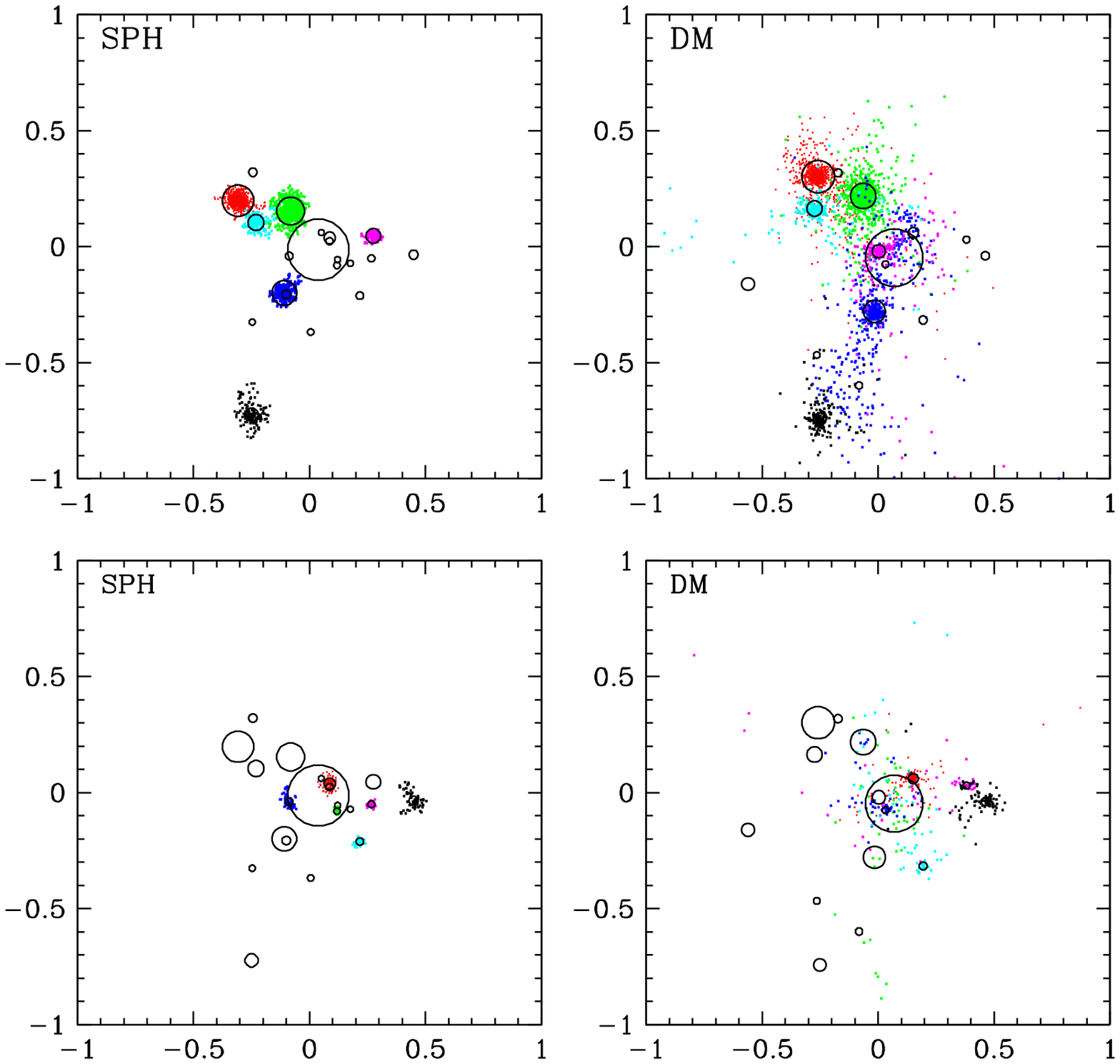}
}
\caption{
Correspondence of particles in subhalos of the 
fourth largest halo of the
SPH simulation
(left panels) and the DM simulation (right panels).  Circles
show subhalos of the two simulations as in the bottom
panels of Fig.~\ref{fig:circle}.
Colored dots show particles associated with particular subhalos
in the SPH simulation (left) and the locations of the corresponding
particles in the DM simulation (right).  In the upper panels,
the six most massive subhalos (after the most massive, central
subhalo) are marked, while the lower panels show lower mass
subhalos that illustrate a range of behaviors.
}
\label{fig:halomark}
\end{figure}

\begin{figure}
\centerline{
\epsfxsize=5.5truein
\epsfbox[55 180 560 725]{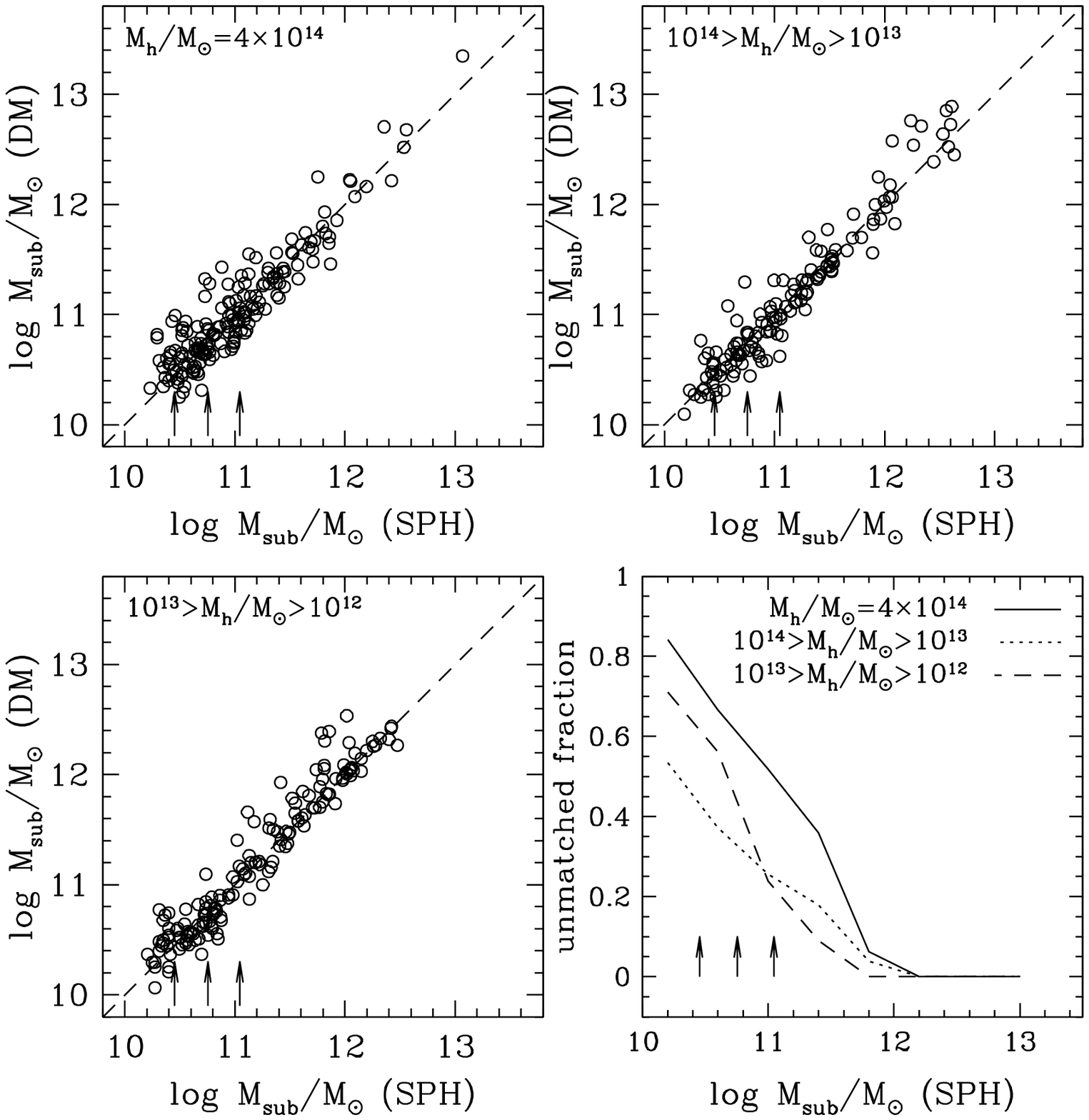}
}
\caption{
Masses of subhalos in the DM simulation vs. masses of the corresponding
halos in the SPH simulation.  Different panels show different host
{\it halo} mass ranges as indicated.  A DM subhalo is identified as
corresponding to an SPH subhalo if it contains more than 1/4 of the
same particles {\it and} is not more than a factor of four more massive;
the latter condition eliminates spurious matches of tidally stripped
particles to much larger subhalos.  The lower right panel shows the
fraction of SPH halos that have no matching DM halo by these criteria.
Vertical arrows in each panel mark the masses corresponding to 32, 64,
and 128 particles.  
}
\label{fig:masscomp}
\end{figure}

Figures~\ref{fig:halomark} and~\ref{fig:masscomp} show that
the correspondence of subhalos remains good, though not perfect,
even at fairly low masses.  Figure~\ref{fig:halomark} 
focuses on the fourth most massive halo, also shown in the
bottom row of Figure~\ref{fig:circle}.  In the upper left panel,
small dots show the particles in six of the seven largest SPH
subhalos, with a different color for each; 
we have omitted the largest subhalo to preserve visual clarity.
Dots in the upper right panel represent the corresponding particles
in the DM simulation (i.e., those that had the same positions in
the initial conditions), with the same color coding.
In every case, there is a clear identification between an SPH subhalo
and a DM subhalo, but particles in the DM simulation are more spread out.
Since the subhalos are identified in the SPH simulation, it is virtually
inevitable that the particle distributions will be more compact
there, but the blue-dot halo, in particular, shows signs of
substantial tidal stripping in the DM simulation.
Of particular interest in this comparison is the magenta-dot
subhalo, which is at a significantly different location in
the DM simulation but has much the same particle content, though
it, too, shows signs of some tidal stripping.  

The lower panels of Figure~\ref{fig:halomark} show similar results
for six of the lower mass subhalos, ranging from 50 particles (black points)
to 156 particles (red points).
The black-, red-, and magenta-dot subhalos have maintained their identity
in the DM simulation, though the latter two have shifted positions
noticeably.  The cyan-dot halo retains a core of the same particles
at about the same location in the DM simulation, but many of its
particles have been tidally stripped and are spread throughout
the core of the halo.  Finally, the blue-dot and green-dot halos appear to 
have been tidally disrupted, with their particle contents widely
dispersed through the halo in the DM simulation.

Figure~\ref{fig:masscomp} shows quantitative results for the full
halo population.  For each SPH subhalo, we match a DM subhalo if
it contains more than 1/4 of the same particles.  We suppress matches
in which the DM subhalo is more than four times the mass of the SPH 
subhalo, since these cases arise when tidally stripped particles
are attached to a different subhalo (typically the central one).
The first three panels compare the masses of DM subhalos to masses
of the matched SPH subhalos, in the 
$4\times 10^{14}M_\odot$ halo (upper left), the halos with
$10^{13}M_\odot < M < 10^{14}M_\odot$ (upper right), and 
$10^{12}M_\odot < M < 10^{13}M_\odot$ (lower left).
The agreement in subhalo masses is generally very good, with
somewhat larger scatter for the least massive subhalos in the
most massive halo.  The lower right panel shows the fraction of
subhalos that are unmatched as a function of subhalo mass.
For the most massive halo, this fraction rises to 50\% for
$M_{\rm sub} \sim 10^{11} M_\odot$, corresponding to $\sim 128$
particles (rightmost vertical arrow).  For the lower mass
halos, the matched fraction is still $\sim 75\%$ at this
$M_{\rm sub}$, and it does not fall to 50\% until $\sim 64$
particles ($10^{13}M_\odot < M < 10^{14}M_\odot$) or $\sim 25$ particles
($10^{12}M_\odot < M < 10^{13}M_\odot$).
It is not clear whether the ``missing'' subhalo matches are primarily
a consequence of physical disruption at low masses or numerical 
artifacts at low particle number, but our results suggest that one
should be cautious in interpreting subhalo mass functions in 
cluster simulations below $\sim 100$ particles or $\sim 10^{11}M_\odot$.

\begin{figure}
\centerline{
\epsfxsize=3.5truein
\epsfbox[100 415 460 720]{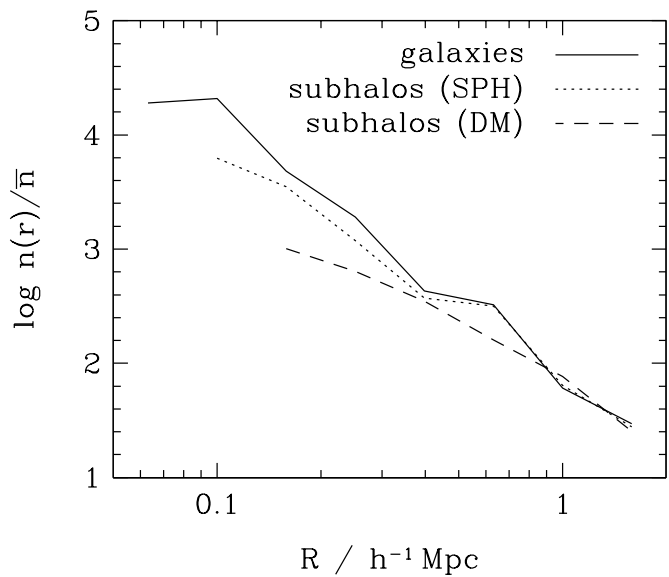}
}
\caption{
Radial number density profile of galaxies (solid line) and subhalos (dotted and
dashed lines for the SPH and DM simulations, respectively)
around the most massive galaxy of the most massive halo.
Galaxies are selected above a mass threshold 
$M_{g,{\rm min}}=7.1\times 10^{9} M_\odot$, and subhalos
are selected above a mass threshold $M_{g,{\rm min}}\Omega_{\rm dm}/\Omega_b$.
Densities are normalized to the mean density of galaxies or subhalos
above the mass threshold in the full simulation volume.  
Curves stop when the only interior
galaxy or subhalo is the central object.
}
\label{fig:profile}
\end{figure}

Returning to the top row of Figure~\ref{fig:circle}, one can see
in the densest regions of the largest halo a
slight paucity of subhalos (relative to the galaxies)
in the SPH simulation and a 
more substantial lack of subhalos in the DM simulation.
Figure~\ref{fig:profile} compares the radial number density profiles of
galaxies and subhalos around the central galaxy of the 
main component of this halo.  Here we use a mass threshold
of $7.1\times 10^9 M_\odot$ for the galaxies and a
mass threshold larger by $\Omega_{\rm dm}/\Omega_b$ for
the subhalos, where $\Omega_{\rm dm}=\Omega_m-\Omega_b$.
Since we normalize the profiles to the mean density of
the corresponding galaxy or subhalo population in the entire simulation volume,
the qualitative appearance of Figure~\ref{fig:profile} is not
sensitive to the choice of mass thresholds.
The radial profile of the SPH
subhalos is only slightly depressed relative to the galaxies,
but the DM subhalos are substantially depleted within 
$R \sim 0.2\hmpc$.  This result agrees with other recent
studies of subhalo depletion in cluster mass halos
\citep{diemand04,gao04,nagai05},
although the impact of the baryons on the survival
of subhalos appears somewhat stronger here than in
\cite{nagai05}.

Can one use
substructure in a high-resolution N-body simulation to identify the
galaxy population that would be found in a full hydrodynamic simulation?
Here we will focus on galaxy populations defined by thresholds in
baryonic mass (stars plus cold, dense gas), which should be similar
but not identical to populations defined by thresholds in luminosity.
Two subtleties then arise in trying to answer the question.
First, since halos retain an enormous amount of substructure
if one goes to sufficiently small mass scales 
(e.g., \citealt{moore99,springel01}), it is virtually guaranteed that
one can find ``enough'' substructures in each halo to correspond to
the number of galaxies above a moderate or high baryonic mass threshold.
However, the N-body/substructure approach has no predictive power unless 
one knows {\it which} (or at least how many) substructures to pick
in each halo, so there must be some threshold in substructure mass
for a given baryonic mass threshold.

The second subtlety arises because a halo that falls into a larger halo
(and thus becomes a  substructure) starts to lose mass via tidal stripping.
Since this process does not remove mass from the halo's central galaxy
(at least until it is close to total disruption), the mass of a substructure
containing a galaxy in a group or cluster will generally be smaller
than the mass of an isolated halo that contains a similar galaxy in
the ``field.''  It is therefore unlikely that a simple global threshold
in substructure mass is likely to work for identifying a galaxy population ---
if one picks the threshold based on the lowest mass halos that host such
galaxies in the field, then there will be too few ``galaxies'' found
in rich groups and clusters.  One way to tackle this problem is to
use circular velocity thresholds instead of mass thresholds, in the
hope that the circular velocity remains a nearly monotonic function
of the central galaxy's baryonic mass even if tidal stripping removes
the outer parts of the halo in which the galaxy formed.  
This approach suffers from ambiguity in the choice of where
to define the circular velocity, especially since tidal stripping 
alters the density profiles of substructures, making them systematically
different from isolated halos \citep{stoehr02}.
Furthermore, recent N-body studies indicate that the
circular velocities of subhalos do in fact decline as they 
are tidally stripped \cite{nagai05}, so using circular
velocity instead of subhalo mass only partly compensates for
stripping effects.

\begin{figure}
\centerline{
\epsfxsize=5.5truein
\epsfbox[50 490 530 720]{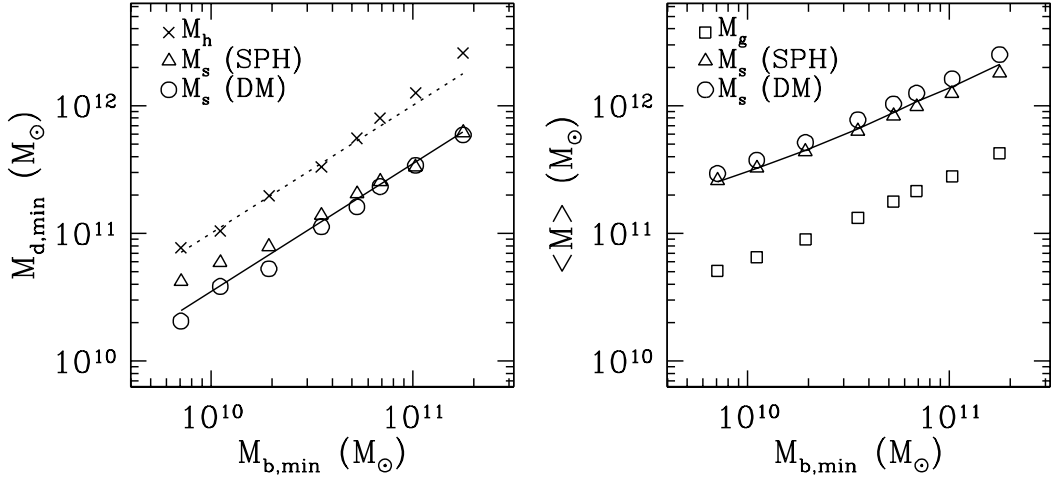}
}
\caption{
(Left) Mass thresholds for identifying dark matter subhalos with SPH
galaxies.  Crosses show the halo mass $M$ at which $\Navg=0.5$ for
galaxies above the threshold baryonic mass $M_{b,{\rm min}}$.
When matching a subhalo population to a galaxy population, we only
consider {\it halos} above this threshold mass.  Triangles and 
circles show the {\it subhalo} mass threshold that is then required
to match the space density of the galaxy population, in the SPH and
DM simulations, respectively.  For comparison, the dotted line shows
$1.35\times \Omega_{\rm dm}/\Omega_b \times M_{b,{\rm min}}$, 
and the solid line shows $3.5 M_{b,{\rm min}}$.
(Right) The average mass of galaxies (squares) above $M_{b,{\rm min}}$
compared to the average mass of dark matter subhalos (triangles and
circles for the SPH and DM simulations, respectively) 
above the mass threshold indicated in the left panel.
The solid line shows the mean galaxy mass multiplied by five.
}
\label{fig:thresholds}
\end{figure}

Here we have adopted a simple approach that seems to work surprisingly well.
To identify a substructure population that corresponds to the galaxy
population above mass threshold $\Mbmin$, we first apply a global mass
threshold to the {\it halo} (not substructure) population.
In the SPH simulation, we find the halo mass $\Mhmin$ at which 50\%
of halos (in a sample of 20 centered on that mass)
contain a galaxy above $\Mbmin$, and we eliminate halos with
$M< \Mhmin$.  We use the same threshold mass in the DM simulation,
except that we multiply $\Mhmin$ by $\Omega_m/(\Omega_m-\Omega_b)$
to account for the fact that baryons are not counted when computing
the mass of the SPH halos.  In the left hand panel of 
Figure~\ref{fig:thresholds}, crosses show the minimum halo mass
(in the SPH simulation) as a function of the galaxy mass threshold.
These lie close to the line $1.35(\Omega_m-\Omega_b)/\Omega_b$,
indicating that these minimum mass halos have typically put
about 75\% of their available baryons into the central SPH galaxy.

After eliminating halos below the mass threshold,
we now apply a global mass threshold $\Msmin$ to the {\it subhalo} population,
choosing its value so that the total number of subhalos in
the simulation is equal to the total number of galaxies above the
mass threshold.  These subhalos are the ``galaxy'' population
predicted by the subhalo method.
The $\Msmin$ threshold is always lower than $\Mhmin$ because
of the tidal stripping effects discussed above.
Note also that a halo that passes the $\Mhmin$ threshold may not,
in the end, contain a galaxy, since the mass of its largest substructure
may be lower than $\Msmin$.  These ``unoccupied'' halos (which are, in
practice, fairly rare) should represent cases where the
halo contains two or more galaxies below the $\Mbmin$ threshold instead
of one (or more) above it.

Triangles and circles in 
the left panel of 
Figure~\ref{fig:thresholds} show the threshold values required
to match the galaxy and subhalo populations, for the SPH and DM
simulations, respectively.  
The threshold for the DM simulation is approximately 3.5 times
the galaxy baryonic mass, at all masses.
The threshold for the SPH simulation is higher than this at
low galaxy masses, suggesting that in this regime the baryonic
clumps help to reduce tidal mass loss from their local subhalos.
The right panel of Figure~\ref{fig:thresholds}
shows the average masses of substructures above these thresholds.
To a good approximation, the mean mass of subhalos above a threshold
is simply five times the mean mass of galaxies above the corresponding
threshold.  
We will show below that the halo occupation statistics and spatial
clustering of the subhalo populations identified in this way
are similar to those of the corresponding SPH galaxy populations.
However, we first investigate the extent to which the presence
of dissipative baryons in the SPH simulations alters the properties
of the dark matter subhalos themselves.

\subsection{The Influence of Baryons on Halo Substructure}
\label{sec:influence}

\begin{figure}
\centerline{
\epsfxsize=5.5truein
\epsfbox[40 210 560 720]{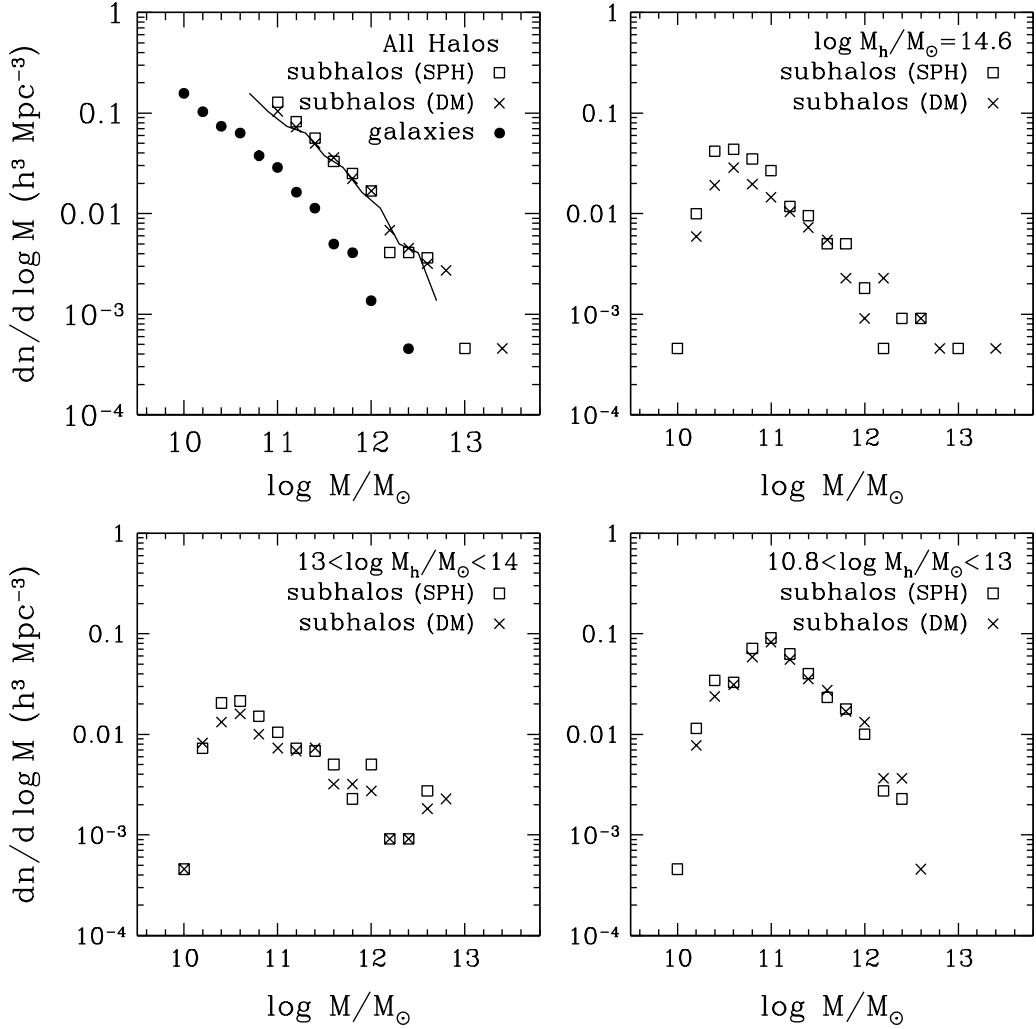}
}
\caption{
Differential mass functions, in number per $(\hmpc)^3$ per decade,
of galaxies and subhalos.  The upper left panel shows results for the
full simulation, with filled circles showing the baryonic mass
function of SPH galaxies and squares and crosses showing the 
dark matter mass function of subhalos in the SPH and DM simulations,
respectively.  Here subhalos have been selected according to the
two-stage thresholding procedure described in the text.  
The solid line shows the galaxy mass function shifted right by
a factor of five.  Note that bins are evenly spaced in $\log M$
but that some bins at high mass contain no objects.
Remaining panels show the subhalo mass functions in bins of halo
mass, as indicated.  
The lower limit of the $10.8 < \log M_h < 13$ bin corresponds to a mass
of 64 dark matter particles.  For these three panels we include all
identified substructures, not just those selected by the thresholding
procedure.
}
\label{fig:massfun}
\end{figure}

Figure~\ref{fig:massfun}a compares the differential baryonic
mass function of SPH galaxies to the differential mass functions
of subhalos in the SPH and DM simulations.
The two subhalo mass functions are similar, showing that the
dissipative baryon component has only a small impact on this
global measure of the subhalo population.
The subhalo mass functions are similar in form to the
galaxy mass function, shifted in mass scale by a factor of five,
with a somewhat larger shift at low masses.

The remaining panels show the subhalo mass function in the largest
halo (Fig.~\ref{fig:massfun}b) and in halo mass ranges
$\log M_h/M_\odot = 13-14$ (Fig.~\ref{fig:massfun}c) and
$10.8-13$ (Fig.~\ref{fig:massfun}d).  Here we have not imposed
any explicit threshold on the subhalo masses.
The turnover of the subhalo mass functions at low masses is an
artifact of the simulations' finite mass resolution, 
but since this resolution is the same in each case, we 
can use the differential comparison to investigate the
influence of the baryon component on the survival of dark matter subhalos.
These effects are generally mild, but they have the expected sign.
In particular, the ability of dense baryon clumps to retain surrounding
dark matter concentrations boosts the number of low mass subhalos in
the largest halos, by up to a factor of two.

\begin{figure}
\centerline{
\epsfxsize=3.5truein
\epsfbox[50 470 320 730]{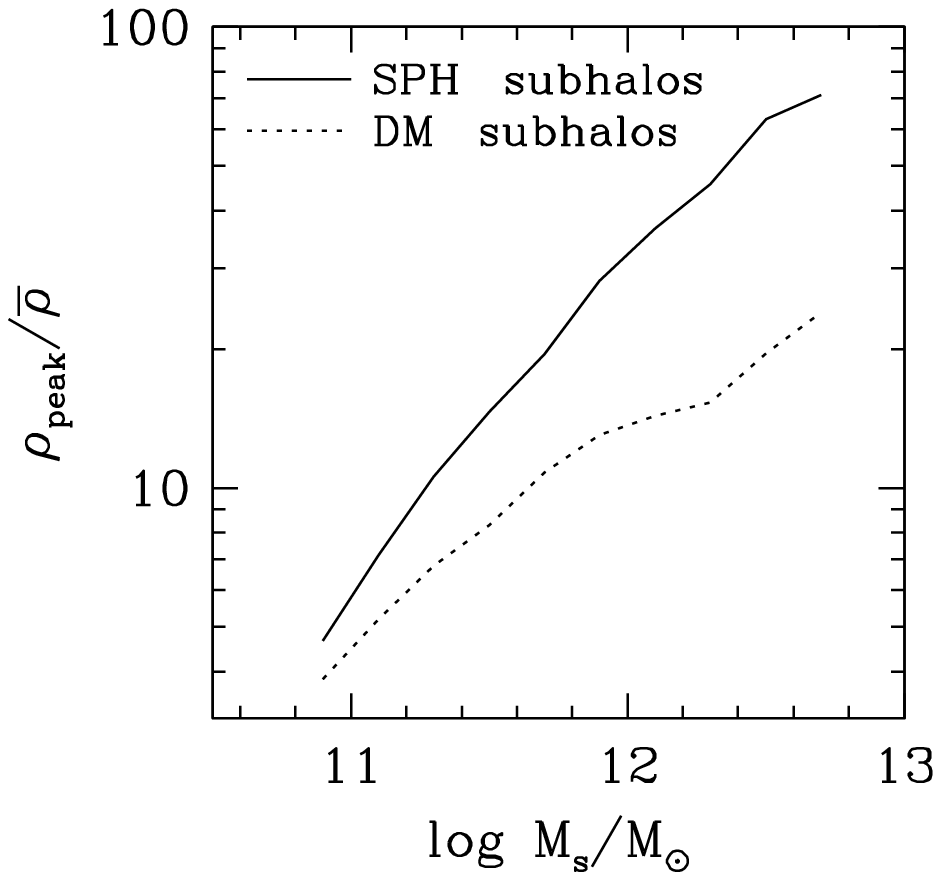}
}
\caption{
Average peak-to-mean density ratio of dark matter subhalos in the SPH 
simulation (solid line) and DM simulation (dotted line).
Densities are computed for each particle in the
subhalo using an SPH-like smoothing kernel containing 32 neighbors,
and the quantity shown is the ratio of the maximum density in the subhalo
to the mean density of the subhalo, averaged in bins of subhalo mass.
The trend with subhalo mass is probably a mass resolution effect,
but the difference between the SPH and DM simulations shows the concentrating
effect of cooled baryonic components within subhalos.
}
\label{fig:peak}
\end{figure}

Figure~\ref{fig:peak} examines the influence of baryons on the internal
structure of subhalos.  We define a simple measure of subhalo ``concentration''
by measuring a density for each particle using an SPH-like smoothing
kernel containing 32 neighbors and taking the ratio of the highest density
in the subhalo to the mean density of all particles in the subhalo.
The strong trend of this concentration measure with subhalo mass
is probably an effect of mass resolution --- one can trace the density profile
of more massive subhalos into smaller radii --- but we can again
make a differential comparison.
In contrast to Figure~\ref{fig:massfun}, the influence of baryons 
on subhalo concentrations is strong, with differences of a factor
of several by this measure.  This result is in qualitative agreement
with the expectation that dissipative baryons lead to adiabatic
contraction of their surrounding dark matter concentrations
\citep{blumenthal86}, though a quantitative investigation of
these effects is better carried out with higher resolution simulations
of individual halos (e.g., \citealt{gnedin04}).

\subsection{Halo Occupation Statistics}
\label{sec:hod}

On scales that are large compared to the virial diameters of the
largest halos, the clustering of the galaxy population is determined
by the number of galaxies in each halo, regardless of their internal
distribution within halos.  If the halo occupation statistics of
a population of galaxies and a population of subhalos are identical,
then they will yield the same large scale results for all measures
of clustering.  Here we compare halo occupation statistics for
mass-thresholded samples of SPH galaxies to those of matched subhalo
samples defined by the two-stage thresholding procedure 
described in \S\ref{sec:populations}.  

\begin{figure}
\centerline{
\epsfxsize=5.5truein
\epsfbox[30 50 520 720]{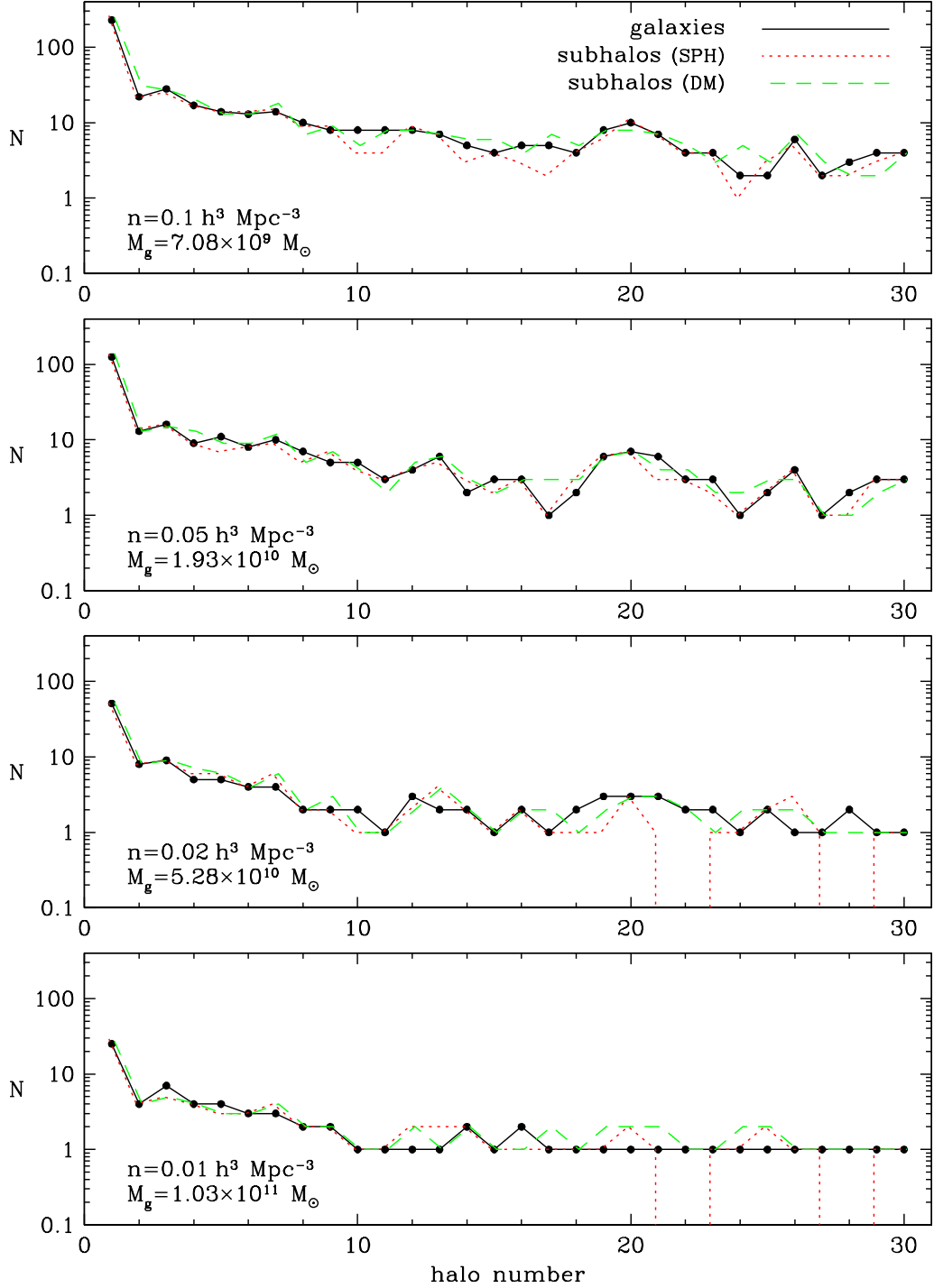}
}
\caption{
Halo occupations of the 30 most massive halos in the SPH and DM simulations.
Points connected by the solid black line represent galaxies, while the
dotted red and dashed green lines represent the matched subhalo populations
in the SPH and DM simulations, respectively.  From top to bottom, the
four panels correspond to increasing galaxy baryonic mass thresholds
and decreasing population space densities, as indicated in each panel.
}
\label{fig:nghalo}
\end{figure}

Points in Figure~\ref{fig:nghalo} show the number of galaxies
in each of the 30 most massive halos of the SPH simulation.
The four panels correspond to four different baryonic mass thresholds,
and the mean space densities of galaxies above these thresholds
are 0.1, 0.05, 0.02, and $0.01\vunits$, respectively.
Dotted lines show the number of subhalos in each of these halos
above the mass thresholds indicated by the triangles in
Figure~\ref{fig:thresholds}a; by construction, the mean space
density of these subhalos matches that of the corresponding
galaxy population.  Dashed lines show the number of subhalos in
the same halos of the DM simulation, with mass thresholds shown
by the circles in Figure~\ref{fig:thresholds}a.
The agreement between the number of galaxies and the number of
subhalos in the matched population is extraordinarily good for
both the SPH and DM simulations, at all four space densities.
This agreement holds for the most massive halo despite the visible paucity
of subhalos in the densest regions of this halo (Fig.~\ref{fig:circle}).
The subhalo mass threshold, chosen to give agreement with the global
number density of galaxies in the simulation, has the effect of
replacing these missing subhalos in the halo core with slightly
less massive subhalos in the outskirts.

\begin{figure}
\centerline{
\epsfxsize=4.5truein
\epsfbox[40 50 530 720]{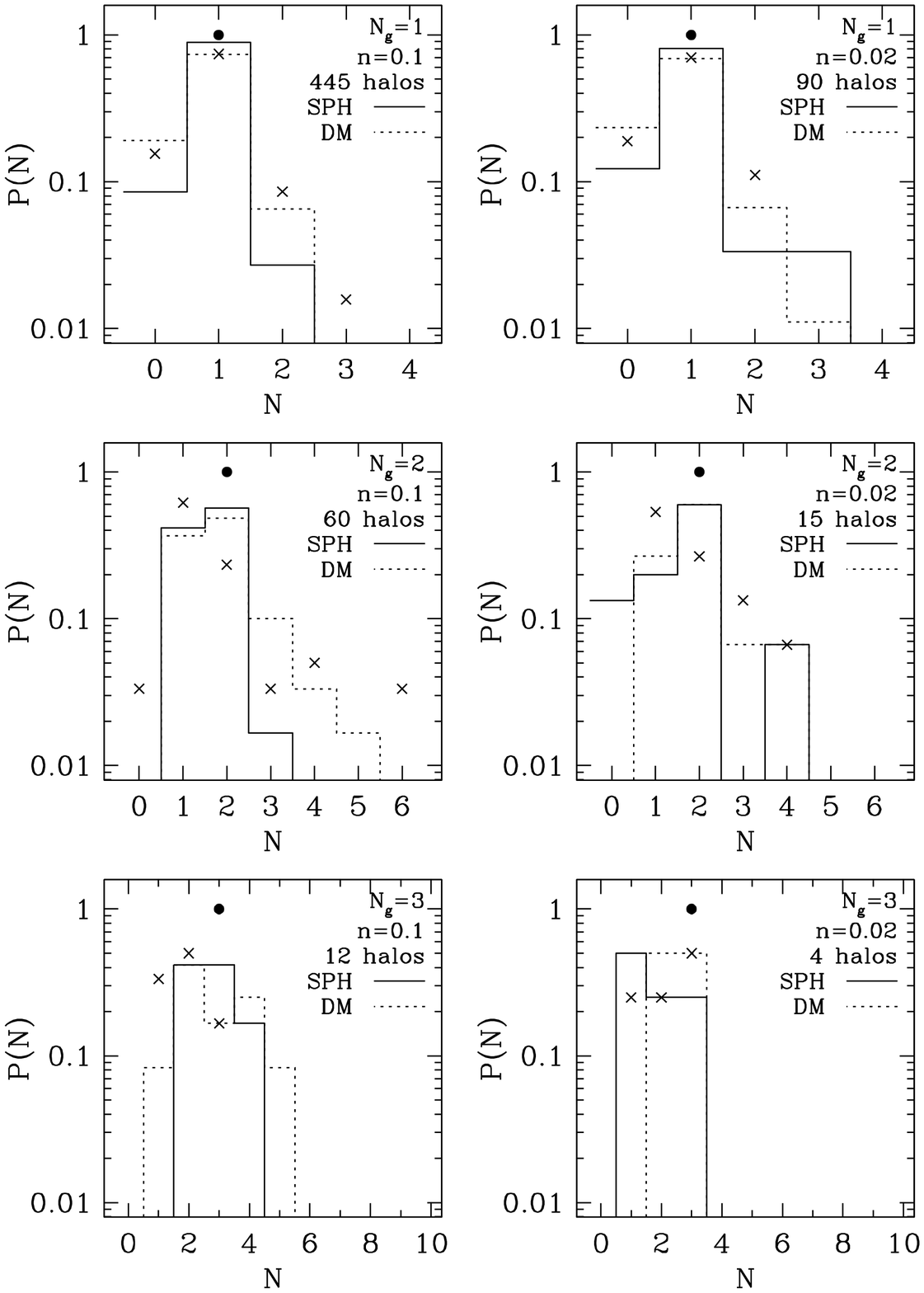}
}
\caption{
Distribution of subhalo occupations in halos with low galaxy occupation
number.  Top, middle, and bottom panels show results for halos that
contain 1, 2, and 3 SPH galaxies, respectively, in the $n=0.1\vunits$
sample (left) and the $n=0.02\vunits$ sample (right).  The corresponding
halo masses can be inferred from Fig.~\ref{fig:navg}.
Solid histograms show the distribution of the number of subhalos above
the matching mass threshold in the SPH simulation, and dotted histograms
show the same distribution for the DM simulation.  Perfect 
agreement would correspond to $P(N)=1$ for $N=N_g$ and $P(N)=0$ otherwise,
in which case the histograms would intersect the filled circles.
Crosses show $P(N)$ for a ``control'' sample in which the galaxy occupation
of each halo is replaced by the galaxy occupation of the next halo in
a list rank-ordered by mass.
}
\label{fig:pn}
\end{figure}

Figure~\ref{fig:pn} examines the distribution of subhalo numbers
in halos with $N_g=1$, 2, or 3 SPH galaxies (top to bottom),
for samples with space densities of $0.1\vunits$ (left hand panels)
or $0.02\vunits$ (right hand panels).
In each panel, solid histograms show the distribution of subhalo
numbers in the SPH simulation, and dotted histograms show the
distribution in the DM simulation.
If the subhalo and galaxy populations agreed perfectly, these
histograms would be $P(N)=1$ for $N=N_g$ and $P(N)=0$ for all
other $N$.  The agreement is generally very good but not perfect.
For example, of the 15 halos with $N_g=2$ for $n=0.02\vunits$,
nine have two subhalos in the DM simulation, four have one subhalo,
one has three, and one has four.  The agreement for the SPH subhalos
is usually better than for the DM subhalos, but not dramatically so.

The detailed agreement seen in the individual halo plots of
Figure~\ref{fig:circle} and in the number counts of Figure~\ref{fig:nghalo}
suggests that the subhalo counts are not just reproducing the
mean number of galaxies at a given halo mass but are, to some degree,
tracking the variation in galaxy number from halo to halo at each mass.
To investigate this issue in the low occupancy regime,
we rank-ordered the halo list by mass, then
replaced each halo's galaxy occupation $N_g$ with the occupation $N_g^\prime$
of the next halo on the list.
Crosses in Figure~\ref{fig:pn} show $P(N_g^\prime)$ --- in essence,
they show the effect of randomly replacing each halo's galaxy population
with that of another halo of nearly identical mass.
The subhalo method is doing ``better than random'' if the histogram
lies above the cross in the $N=N_g$ bin and below the crosses in the
other bins.  For $N_g=2$, this is clearly the case; one can predict
a halo's galaxy number more accurately using its SPH or DM subhalos
than by using the galaxy number of another halo of similar mass.
For $N_g=1$ and $N_g=3$, on the other hand, subhalos do at most slightly 
better than random assignment.
Note, however, that the absolute level of agreement for $N_g=1$ is
high, and that fewer than 10\% of halos that have two subhalos above
threshold contain only a single galaxy.

\begin{figure}
\centerline{
\epsfxsize=2.5truein
\epsfbox[115 225 515 640]{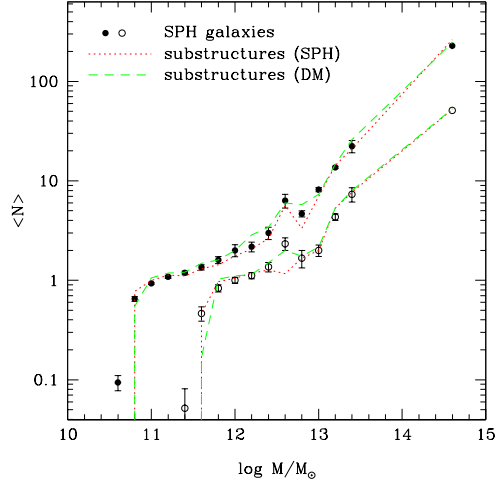}
}
\caption{
The mean number of galaxies (points with error bars) and substructures
(dotted line for the SPH simulation and dashed line for the
DM simulation) in halos of mass $M$.  Filled circles and upper lines
represent the $n=0.1\vunits$ sample, while open circles and lower
lines represent the $n=0.02\vunits$ sample.  Error bars show the
error on the mean, computed from the dispersion of $N$ among all halos
in the bin divided by the square-root of the number of halos.
The highest mass bin contains only a single halo, so no error bar
is computed.
}
\label{fig:navg}
\end{figure}

Figure~\ref{fig:navg} compares the mean occupation functions $\Navg$
of the two galaxy samples to those of the corresponding subhalo samples.
The locations of the lower cutoffs match by construction, since we
eliminate halos below the mass at which $\Navg=0.5$ before choosing
the subhalo population.  In addition, matching the global space density
of the subhalo population to that of the galaxy population forces
agreement in the values of $\int_0^\infty dn/dM \Navg dM$, where
$dn/dM$ is the halo mass function.  However, it is clear that the
agreement between the galaxy and subhalo occupation functions is
far better than these constraints alone would require. 
The excellent match in halo occupations seen here and in 
Figure~\ref{fig:nghalo} implies that the large scale clustering
of a mass-thresholded galaxy population and a properly matched
subhalo population should be very similar in all respects.

\begin{figure}
\centerline{
\epsfxsize=2.5truein
\epsfbox[115 225 515 640]{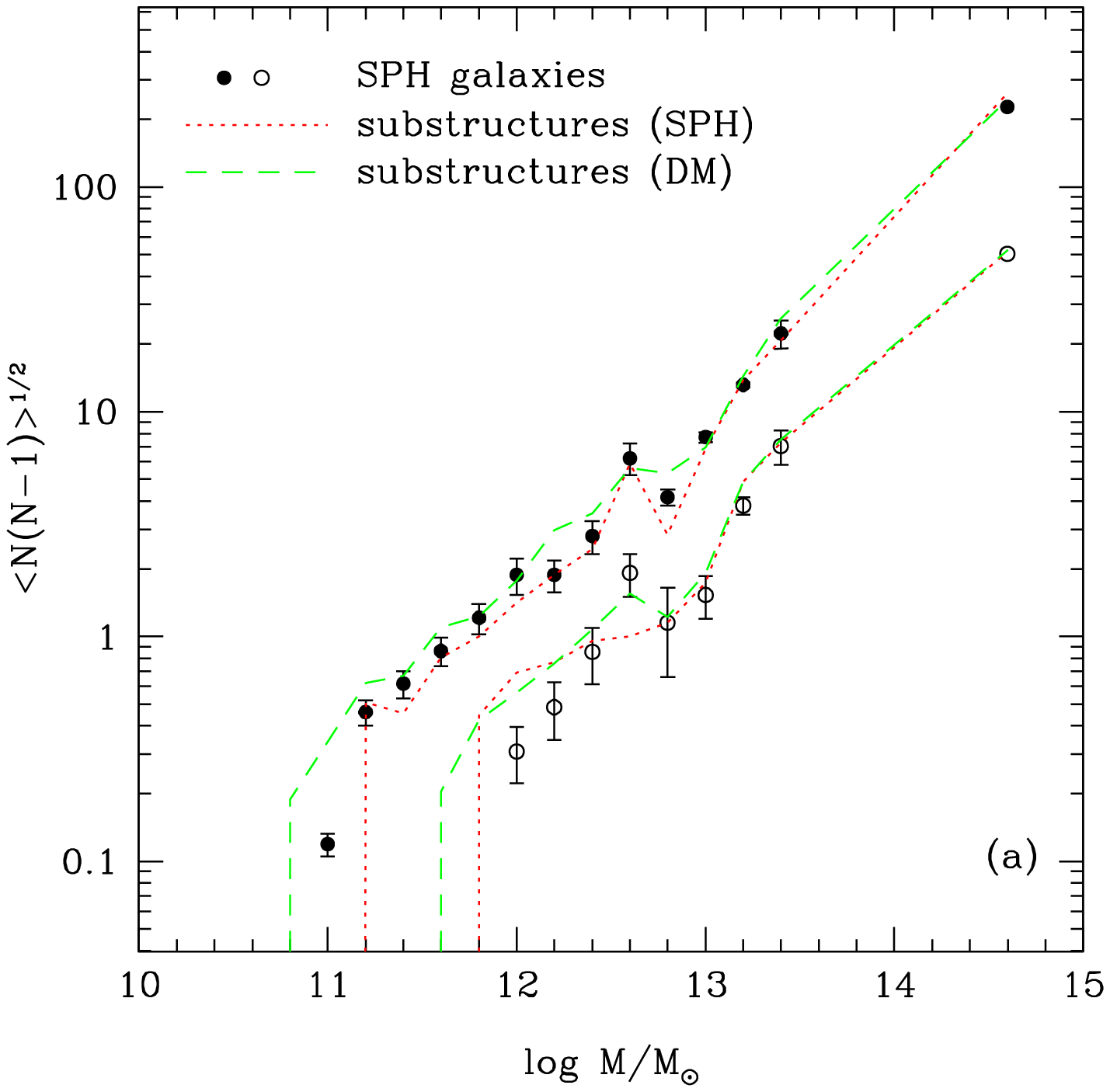}
\hskip 0.25truein
\epsfxsize=2.5truein
\epsfbox[115 225 515 640]{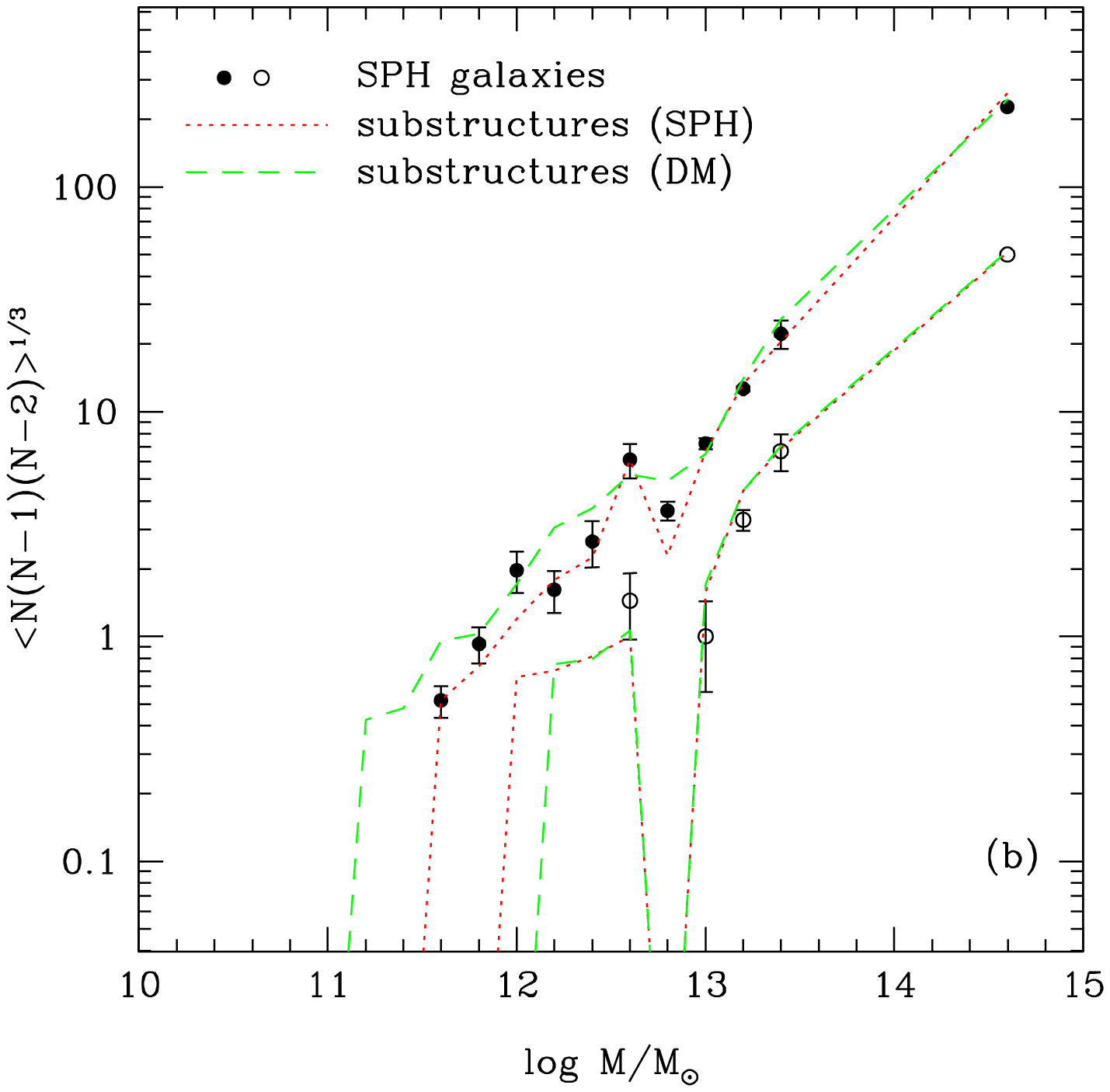}
}
\caption{
{\it (Left)} The 
square-root of the mean number of pairs per halo, $\NNavg^{1/2}$,
in the same format as Fig.~\ref{fig:navg}.
{\it (Right)} The cube-root of the mean number of triples per halo,
$\NNNavg^{1/3}$.  If $P(N|\Navg)$ were Poisson distributed,
then both of these quantities would trace the mean occupation
functions $\Navg$, but comparison to Fig.~\ref{fig:navg} shows
that the distributions for all three populations are substantially
narrower than Poisson distributions at low occupation numbers.
}
\label{fig:navg2}
\end{figure}

On small scales, the two-point correlation function is dominated by
pairs of galaxies that reside in the same halo.  In this one-halo
regime, the important quantity for determining the two-point
correlation function is $\NNavg$, the mean number of pairs per halo
\citep{seljak00}.  Similarly, the three-point correlation function
on small scales depends on the mean number of triples per halo $\NNNavg$ 
\citep{scoccimarro01}.  Figures~\ref{fig:navg2}a and~\ref{fig:navg2}b
plot $\NNavg^{1/2}$ and $\NNNavg^{1/3}$, respectively.  Taking square
and cube roots allows direct comparison to the mean occupations
plotted in Figure~\ref{fig:navg}: if $P(N|N_{\rm avg})$ is 
Poisson distributed, then $\NNNavg^{1/3}=\NNavg^{1/2}=\Navg$.
Figure~\ref{fig:navg2} shows good agreement between the pair and
triple counts of galaxy and subhalo populations in all cases.
In particular, the subhalo counts reproduce a key feature of the
galaxy counts, namely a distribution that is substantially narrower than
a Poisson distribution at low occupation numbers.
\cite{berlind03} and \cite{kravtsov04} discuss the physics behind
these sub-Poisson count distributions: over the range in which
$\langle N \rangle$ rises from one to a few, more massive halos tend
to host more massive central galaxies or subhalos instead of hosting multiple
objects above the mass threshold (see also \citealt{benson00,zheng05}).

In the one-halo regime, galaxy clustering depends on the internal 
distribution of galaxies within halos, in addition to $P(N|M)$.
Therefore, agreement in $\NNavg$ and $\NNNavg$ does not guarantee
agreement in two- and three-point correlations on small scales.
We now turn to a direct investigation of small scale clustering
as measured by the two-point correlation function and
moments of counts-in-cells.

\subsection{$\xi(r)$ and $S_n$}
\label{sec:sn}

Because of the small simulation volume, the calculated galaxy clustering
statistics are not good estimates
of the global predictions for this cosmological model.  However, since
the DM and SPH simulations started from identical initial conditions,
we can carry out differential comparisons between the clustering of
SPH galaxies, SPH subhalos, and DM subhalos, and we expect the differences
to be indicative of those that would arise in larger volumes.

\begin{figure}
\centerline{
\epsfxsize=2.5truein
\epsfbox[20 150 570 700]{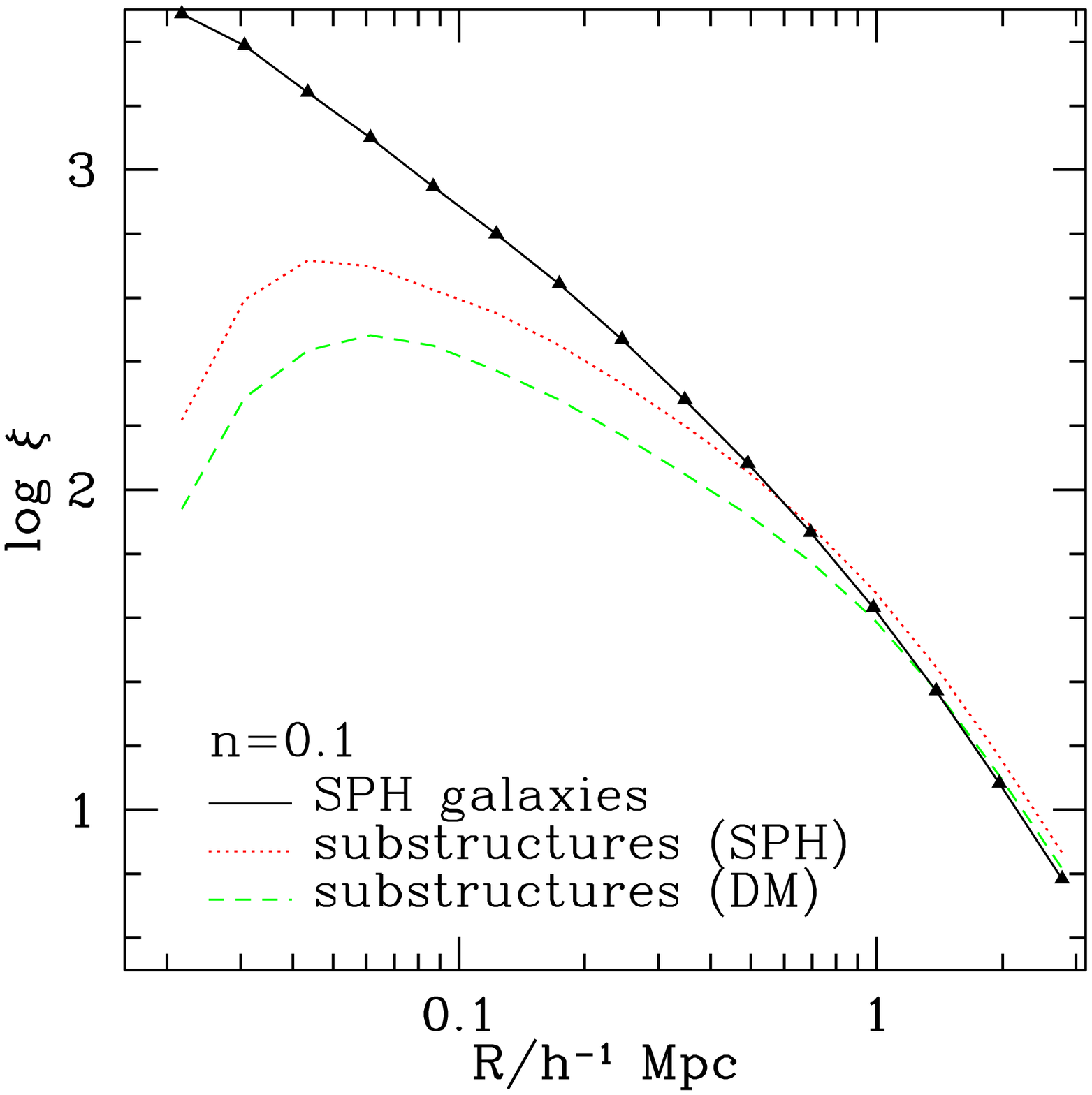}
\hskip 0.25truein
\epsfxsize=2.5truein
\epsfbox[20 150 570 700]{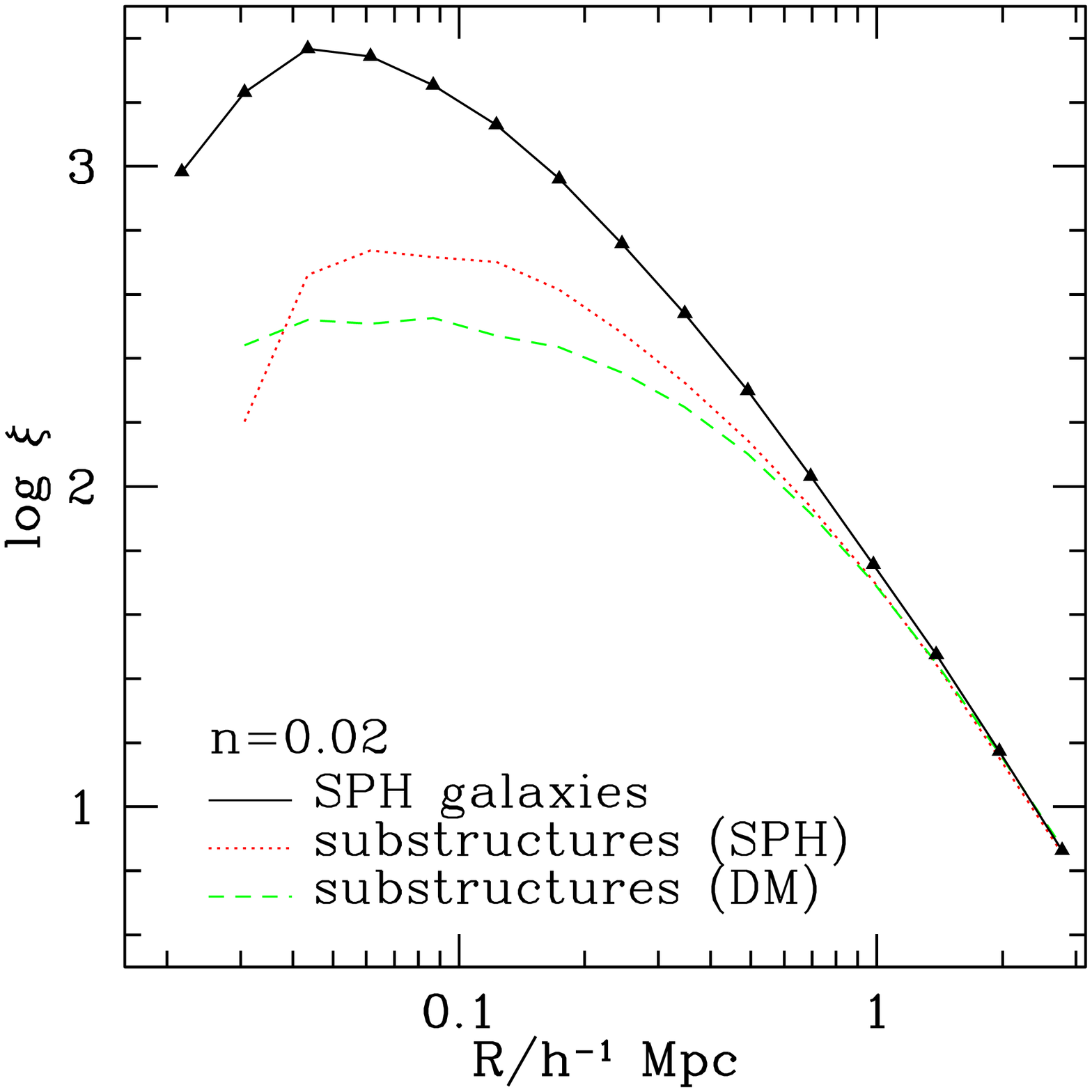}
}
\caption{
Two-point correlation function of
SPH galaxies (solid line),
SPH subhalos (dotted line), and DM subhalos (dashed line).
The left hand panel shows samples with a space density
of $0.1 \vunits$ and the right hand panel shows samples
with higher mass thresholds and a space density of $0.02\vunits$.
}
\label{fig:xi}
\end{figure}

Figure~\ref{fig:xi} shows the two-point correlation functions of
SPH galaxies and subhalos, for samples with a space density of
$0.1\vunits$ (left panel) and $0.02\vunits$ (right panel).
The strong curvature in the range $r \sim 0.1-2\hmpc$, especially
evident for the $0.1\vunits$ sample, is produced by the single
large halo; if we eliminate all members of this halo before 
computing $\xi(r)$, then the correlation functions have an
approximately power-law form.  
At $r \ga 0.5\hmpc$, results for the galaxies, SPH
subhalos, and DM subhalos converge, as expected based on the
similarity of halo occupations shown in \S\ref{sec:hod}.
At smaller scales, the two-point function of subhalos is depressed
relative to that of galaxies, more strongly for the subhalos of 
the DM simulation, and more strongly for the sample with lower
mass threshold ($n=0.1\vunits$).
This departure is primarily caused by the depletion of substructures
in the densest regions of the largest halo, as seen in 
Figure~\ref{fig:circle}.  If we omit this halo, then the
correlation functions of galaxies and subhalos track each other
down to $r \sim 0.2\hmpc$, and only the $n=0.1\vunits$ 
sample shows a clear flattening of the subhalo correlation 
functions at smaller scales.  
\cite{colin99} and \cite{kravtsov04} find subhalo correlation
functions that retain an approximately power-law form down
to $r \sim 0.05\hmpc$.  Since depletion of substructure is 
important mainly in the largest halos, the difference between their 
results and those shown in Figure~\ref{fig:xi} can be explained in
large part by the anomalously large contribution that the largest halo
in our simulation volume makes to $\xi(r)$.
Furthermore, the depletion of subhalos in our $n=0.1\vunits$ sample
could be artificially enhanced by numerical effects, since
the subhalo mass thresholds for this sample correspond to 
only 53 and 23 dark matter particles in the SPH and DM simulations,
respectively.

\begin{figure}
\centerline{
\epsfxsize=2.5truein
\epsfbox[20 150 570 700]{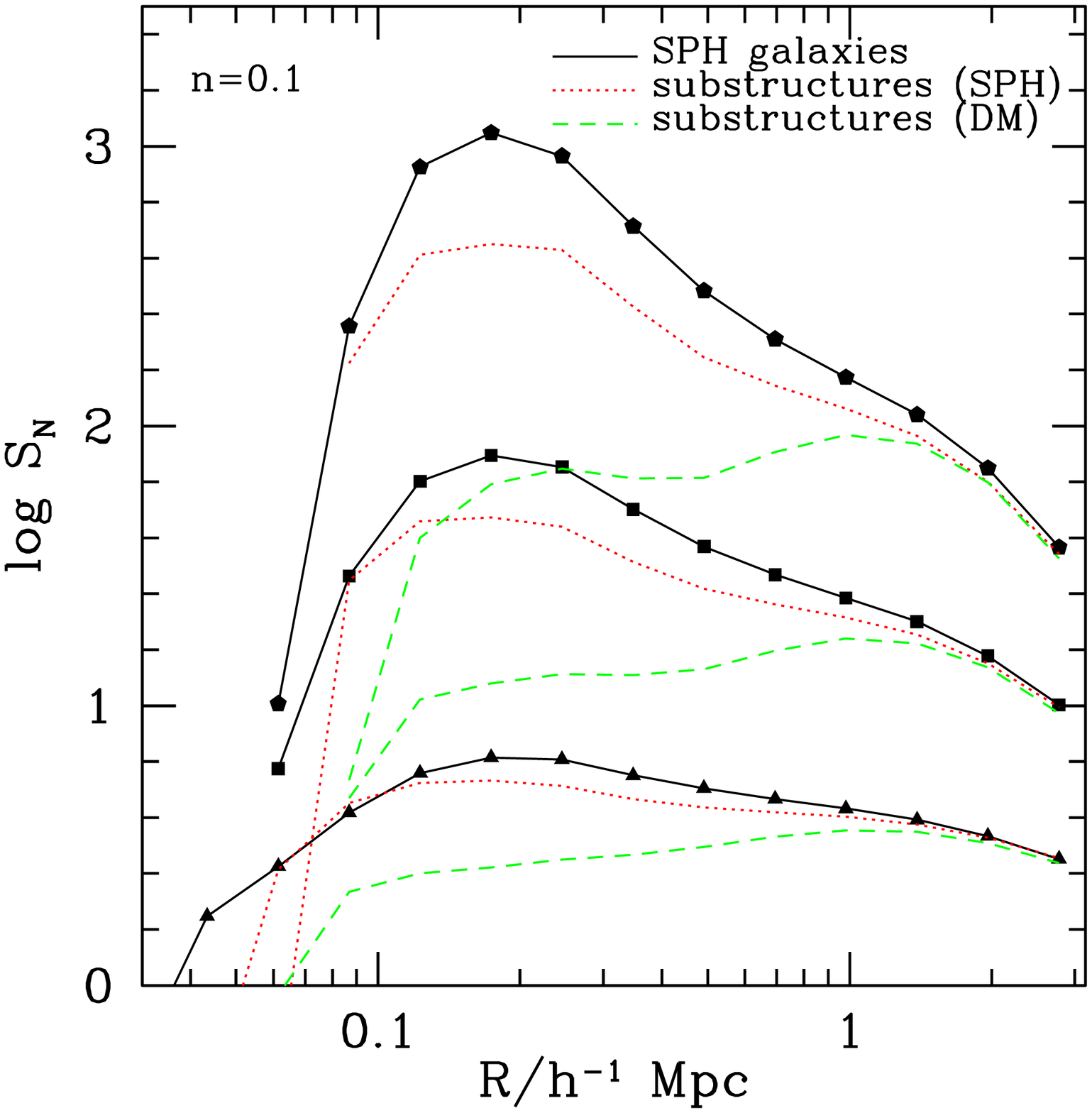}
\hskip 0.25truein
\epsfxsize=2.5truein
\epsfbox[20 150 570 700]{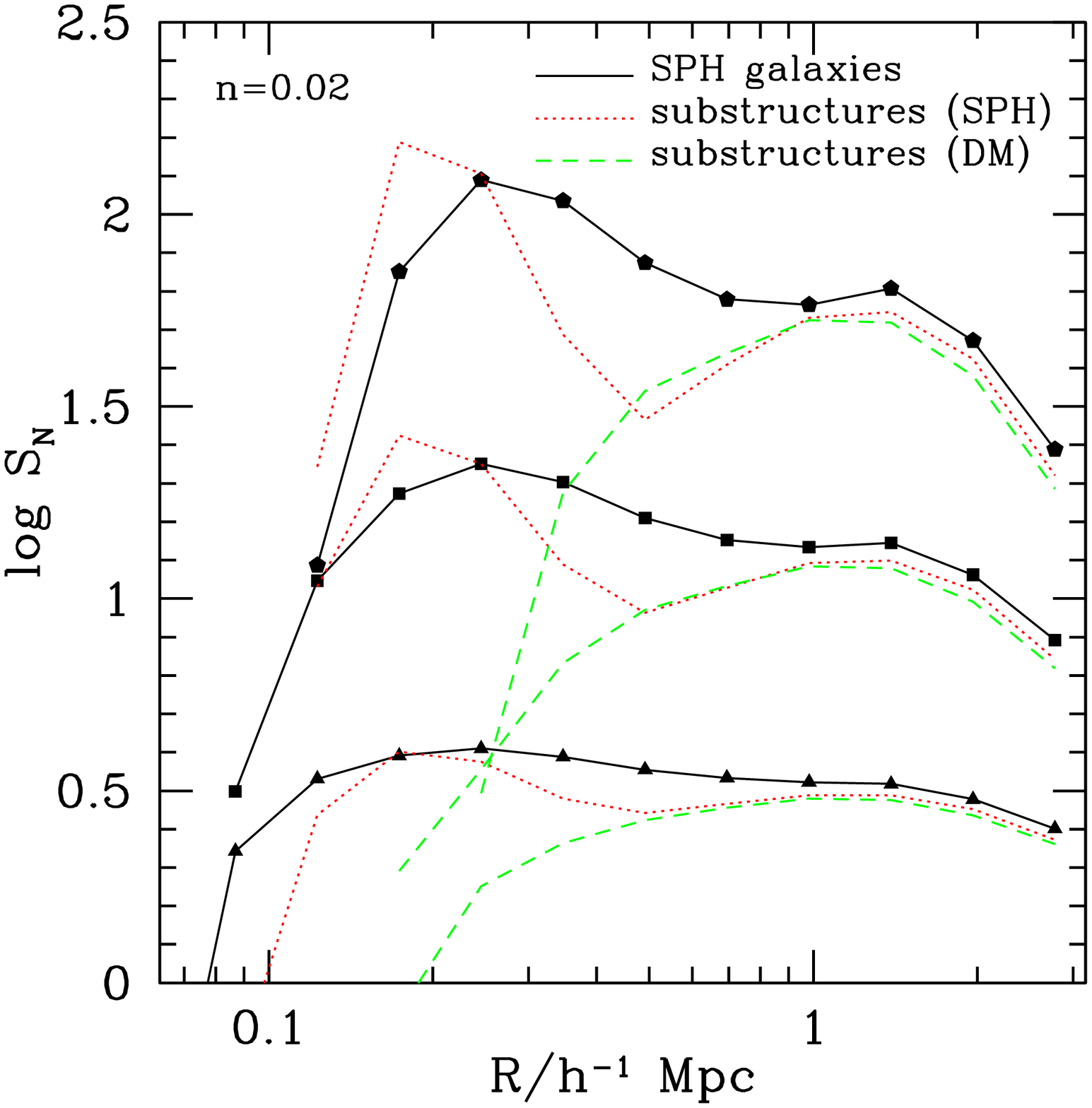}
}
\caption{
Hierarchical moment ratios $S_3$, $S_4$, and $S_5$ (bottom to top)
for SPH galaxies (solid lines),
SPH subhalos (dotted lines), and DM subhalos (dashed lines).
The left hand panel shows samples with a space density
of $0.1 \vunits$, and the right hand panel shows samples
with higher mass thresholds and a space density of $0.02\vunits$.
}
\label{fig:sn}
\end{figure}

As a simple measure of higher order correlations,
Figure~\ref{fig:sn} shows the hierarchical moment ratios
$S_3$, $S_4$, and $S_5$ in spherical cells, estimated from
shot-noise corrected moments of the galaxy and subhalo count
distributions in $2048^3$ cells.
Here $S_n \equiv \langle \delta^n_c \rangle / \langle \delta^2 \rangle^{n-1}$,
where $\langle \delta^n_c \rangle$ is the $n$-th connected moment
of the density field and $\langle \delta^2 \rangle$ is the
variance, both corrected for shot-noise 
(see, e.g., \citealt{szapudi93}).
As with the two-point correlation function,
the hierarchical moment ratios of galaxies and subhalos converge
at scales $R \ga 1\hmpc$, but they differ substantially at
smaller scales, with SPH substructures slightly depressed and
DM substructures more substantially depressed.  
Statistics for
the $n=0.02\vunits$ sample are poor, but
for the $n=0.1\vunits$ sample, the 
depletion of substructure in the densest regions clearly
has an increasing impact for higher moments
of the density field, as one might expect.
The convergence scale is larger for moment ratios (and for the
variance of counts-in-cells) than for $\xi(r)$ because the moments in 
cells of radius $R$ are affected by structure on all scales
smaller than $R$.

\section{Discussion}

Our comparison of matched SPH and N-body simulations illuminates
several aspects of the baryon-dark matter ``conversation.''
For the most part, this conversation is one sided: the dark matter
talks, and the baryons listen.  If one traces the history of
a halo back in time (Figure~\ref{fig:group}), then the population
of the progenitor halos, where gas condenses into galaxies,
is nearly identical in simulations with and without gas.  
At $z=0$, the positions and masses of the larger subhalos in
each halo are similar between the two simulations.
Smaller subhalos have different positions,
presumably because small differences in
the halo potential can modify their orbits, but in many
cases even these lower mass subhalos can be matched based on
particle memberships.  Some SPH subhalos are tidally
stripped or disrupted in the DM simulation.
However, the mass function of subhalos is similar in the two cases,
with baryonic effects producing only a modest enhancement,
primarily for low mass subhalos in high mass halos.
The dissipative baryon component does increase the internal
density of subhalos, and while this has little effect on their
masses in typical environments, it does enhance the survival
of subhalos in the densest regions of massive halos.

These results are not especially surprising, but they are
certainly at the simple end of what might have been expected.
Our most remarkable result is the success of a simple mass
thresholding scheme in identifying subhalo populations 
that have nearly identical halo occupation statistics
to SPH galaxy populations with the same mean space density.
To some degree, the number of subhalos traces the variation
of galaxy number in halos of fixed mass; in particular,
halos with two significant subhalos are more likely to
host two significant galaxies.
For each galaxy mass threshold,
the average mass of host subhalos is approximately five times
the average baryonic mass of the galaxies themselves
(of course, the value of this ratio is likely to depend
on the adopted cosmological parameters).

Our results have encouraging implications for efforts to model
galaxy clustering with the subhalo populations of high resolution,
dissipationless simulations (e.g., \citealt{colin99,kravtsov99,conroy06})
and to develop semi-empirical models of galaxy bias by montonically
matching galaxy luminosity functions to subhalo mass functions
\citep{vale04,vale05}.
The agreement in halo occupation statistics implies that 
SPH galaxies and dark matter subhalos should have similar
large scale clustering statistics, and we indeed find 
good agreement in the two-point correlation function and
the hierarchical moments $S_3$, $S_4$, and $S_5$ on scales
larger than $\sim 1\hmpc$.  However, we find significant
depletion of dark matter substructure in the densest regions
of our one cluster mass halo, in agreement with results
from other groups \citep{diemand04,gao04,nagai05}.
This depletion significantly affects clustering on
scales $\la 1\hmpc$, with galaxy clustering stronger than
SPH subhalo clustering, which in turn is stronger than 
DM subhalo clustering.  The impact on these global statistics
might be exaggerated in our simulations by the dominance
of the largest halo in our small volume, and higher mass
resolution might reduce the depletion effect itself to some degree.
\cite{nagai05} also note that subhalo and galaxy density profiles
are more similar if one selects subhalos based on the mass
they have when they are accreted onto the main halo, rather 
than the final mass, which is preferentially reduced by
tidal stripping in the inner regions.  
\cite{conroy06} show that selecting subhalos based on accreted
mass yields good agreement with observed galaxy clustering
over a wide range of redshifts and luminosities.
Overall,
the present day distribution of dark matter gives one a good
idea of where to place galaxies, and the relation between
subhalo mass and galaxy baryon mass is roughly monotonic
even if one uses the final subhalo mass.
Simple recipes for matching galaxy and subhalo populations
fail in the densest environments, but the large scale
clustering of galaxies is determined mainly by the 
gravitational dynamics of dark matter.

\acknowledgments

We thank Andrey Kravtsov for informative discussions on the
topic of subhalo clustering.
This research was supported by NASA grant NAGS-13308,
NSF grant AST0407125, and the French CNRS.  DHW
thanks the Institut d'Astrophysique de Paris for
generous hospitality during the key phases of this work.



\vfill\eject

\end{document}